\documentclass[11pt]{article}
\usepackage{cospar-umd}
\usepackage{url}

\usepackage{graphicx}
\usepackage[figuresright]{rotating}

\title{MIGRATION OF JUPITER-FAMILY COMETS AND RESONANT ASTEROIDS TO NEAR-EARTH SPACE}

\author{S. I. Ipatov\address{(1) 
George Mason University, USA;
(2) NASA/GSFC, Mail Code 685, 
Greenbelt, MD 20771, USA; e-mail: siipatov@hotmail.com
(for correspondence); 
(3) Institute of Applied Mathematics, 
Moscow, Russia} 
and
J. C. Mather\address{NASA/GSFC, Mail Code 685, Greenbelt,
MD 20771, USA}}

\begin{document}

% typeset front matter
\maketitle

\begin{abstract}

The orbital evolution of about 20000 Jupiter-crossing objects 
and 1500 resonant asteroids under 
the gravitational influence of planets was investigated. 
The rate of their collisions with the 
terrestrial planets was estimated by
computing the probabilities of collisions based on random-phase 
approximations and the orbital elements sampled with a 500 yr step. 
The Bulirsh-Stoer and a symplectic orbit integrators 
gave similar results
for orbital evolution, but sometimes gave different collision
probabilities with the Sun. For orbits close 
to that of Comet 2P, the mean collision probabilities of Jupiter-crossing 
objects with the terrestrial planets were greater by two orders of magnitude 
than for some other comets. For initial orbital elements close to those of 
Comets 2P, 10P, 44P and 113P, a few objects ($\sim$0.1\%) got Earth-crossing orbits 
with semi-major axes $a$$<$2 AU and moved in 
such orbits for more than 1 Myr (up to tens or even hundreds of Myrs). 
Some of them even got inner-Earth orbits (i.e., with aphelion distance $Q$$<$0.983 AU) and Aten orbits. 
Most former trans-Neptunian objects that have typical near-Earth object 
orbits moved in such orbits for millions of years (if they did not disintegrate into mini-comets), so during most of this 
time they were extinct comets. 

\end{abstract}

\section*{INTRODUCTION}

The orbits of more than 70,000 main-belt asteroids, 1000 near-Earth objects (NEOs), 
670 trans-Neptunian objects (TNOs), and 1000 comets are known. Most of the small 
bodies are located in the main asteroid and trans-Neptunian (Edgeworth-Kuiper) 
belts and in the Oort cloud. These belts and the cloud
are considered to be the main sources of the objects that could 
collide with the Earth. 
About  0.4\% of the encounters within 0.2 AU 
of the Earth are from periodic comets 
(http://cfa-www.harvard.edu/iau/lists/CloseApp.html), and 6 out of 20 
recent approaches of comets with the Earth within 0.102 AU  were due 
to periodic comets 
(http://cfa-www.harvard.edu/iau/lists/ClosestComets.html). 
So the fraction of close encounters with the Earth 
due to active comets is $\sim$1\%. 
Reviews of the asteroid and comet hazard were given in [1]-[3]. 
Many scientists [3]-[5] 
%%, e.g. Bottke et al. (2002), Binzel et al. (2002), and Weissman et al. (2002), 
believe that asteroids are the main source of 
NEOs (i.e. objects with perihelion distance $q$$<$1.3 AU). 
Bottke et al. [3] considered that there are 200$\pm$140 km-sized 
Jupiter-family comets 
at $q$$<$1.3 AU, with $\sim$80\% of them being extinct comets.
%%However, Shoemaker {\it et al.} (1994) argued that active and extinct 
%%periodic comets may account altogether for about 20\% of the 
%%production of terrestrial impact craters larger than 20 km in 
%%diameter. There are about 40 active and 800 extinct Earth-crossing 
%%Jupiter-family comets with period $P$$<$20 yr and nuclei $\ge$ 1 km,
%%and about 140--270 active Earth-crossing Halley-family comets 
%%($20<P<200$ yr).

Duncan et al. [6] and Kuchner [7] investigated the migration of TNOs
to Neptune's orbit, and  Levison and Duncan [8] studied the 
migration from Neptune's orbit to Jupiter's orbit. Ipatov and Hahn [9]
considered the migration of 48 Jupiter-crossing objects (JCOs) 
with initial orbits
close to the orbit of Comet P/1996 R2 and found that on average such 
objects spend $\sim$ 5000 yr
in orbits which cross both the orbits of Jupiter and Earth. Using 
these results and additional orbit integrations, and assuming that 
there are  $5\times10^9$ 1-km TNOs with 30$<$$a$$<$50 AU [10], 
%(Jewitt and Fernandez, 2001), Ipatov (1999, 2000, 2001) 
Ipatov [1],[2],[11] found that about $10^4$ 1-km 
former TNOs are Jupiter-crossers now and  10-20\% or more 1-km 
Earth-crossers could have come from the Edgeworth-Kuiper belt into 
Jupiter-crossing orbits. Note that previous estimates of the number 
of bodies with  diameter $d$$\ge$$1$ km and 30$\le$$ a$$ \le$50 AU were 
larger:  $10^{10}$ [12]
%(Jewitt {\it et al.}, 1996) 
and $10^{11}$ [13]
%(Jewitt, 1999). 
In the present paper we use the estimates from [2], 
but now include a much  larger number of JCOs. Preliminary results 
were presented by  Ipatov [14]-[16], who also discussed the 
formation of TNOs and asteroids, and in [17].

%\section*{PROBABILITIES OF COLLISIONS OF NEAR-EARTH OBJECTS WITH PLANETS IN 
\section*{COLLISION PROBABILITIES IN THE MODEL OF FIXED ORBITAL ELEMENTS}

\begin{table}[h]
\begin{center}
\begin{minipage}{14.3cm}

\caption{Characteristic collision times $T_f$ (in Myr)
of minor bodies with planets, coefficient $k$,
and number $N_f$ of simulated objects for the set of NEOs known in 2001.}
$
\begin{array}{lccccc}
\hline
       & $Atens$     & $Apollos$    & $Amors$ &  $NEOs$ & $JFCs$ \\
\hline
$Planet$&T_f~~k~~N_f& T_f~~k~~N_f& T_f~~k~~N_f& T_f~~k~~N_f& T_f~~k\\
$Venus$ & 106~~ 1.2 ~~ 94 & 186~~ 1.7 ~~ 248& -  & 154~~ 1.5 ~~ 343& 
2900~~ 2.5 \\

$Earth$ &15~~ 0.9 ~~110  & 164~~ 1.4 ~~643& 211~~ 2.0~~1 & 67~~ 
1.1~756  &2200~~ 2.3\\

$Mars$  &475~~ 0.4~~6  & 4250~~ 0.9~~574 &5810~~ 1.1~~616 &4710~~ 1.0 
~1197& 17000~~ 1.8 \\
\hline

\end{array}
$
\end{minipage}
\end{center}
\end{table}

In this section we estimate the probabilities of collisions
of near-Earth objects with planets in the model of
fixed orbital elements.
As the actual collisions of migrating objects with terrestrial 
planets are rare, we use an approximation of random phases and 
orientations to estimate probabilities of collision for families of 
objects with similar orbital elements. We suppose that their 
semi-major axes $a$, eccentricities $e$ and inclinations $i$ are 
fixed, but the orientations of the orbits can vary. When a minor body 
collides with a planet at a distance
$R$ from the Sun, the characteristic time to collide, $T_f$, is a 
factor of $k =v/v_c = \sqrt{2a/R -1}$ times that computed with an 
approximation of constant velocity, where $v$ is the velocity at the 
point where the orbit of the body crosses the orbit of the planet, 
and $v_c$ is the velocity for the same semi-major axis and a circular 
orbit.  This coefficient $k$  modifies the formulas obtained by 
Ipatov [1],[18] for characteristic collision and close encounter 
times of two objects moving around the Sun in crossing orbits. These 
formulas also depend  on the synodic period and improve on 
\"Opik's formulas when the semi-major axes of the  objects are close 
to each other. As an example, at $e$=0.7 and $a$=3.06 AU,  we have 
$k$=2.26.

Based on these formulas, we calculated probabilities (1/$T_f$) 
for $\sim$ 1300 
NEOs, including  343 Venus-crossers, 756 Earth-crossers and 1197 
Mars-crossers. The  values of $T_f$ (in Myr), $k$, and the number 
$N_f$ of  objects considered are presented in Table 1. We considered 
separately the Atens, Apollos, Amors, and several Jupiter-family 
comets (JFCs). The relatively small values of $T_f$ for Atens and for 
all NEOs colliding with the Earth are due to several Atens with small 
inclinations discovered during the last three years. If we increase 
the inclination of the Aten object 2000 SG344 from $i$=$0.1^\circ$ to 
$i$=$1^\circ$, then for collisions with the Earth we find $T$=28 Myr 
and $k$=0.84 for Atens and $T$=97 Myr and $k$=1.09 for NEOs.  These 
times are much longer, and illustrate the importance of rare objects. 
%For Jupiter-family comets colliding with the terrestrial planets, 
%$k$$\ge$2. 

%For the three Halley-type comets with periods between 71 and 76~yrs,
%we obtain the characteristic times for the collisions with Earth and Mars
%far exceeding the age of the solar system.
%$T=350$ Gyr and $k=6$ (collision with Earth) and
%$T=3500$ Gyr and  $k=5$ (collision with Mars).

\section*{INITIAL DATA}

\begin{table}
\caption[]{Semi-major axes (in AU), eccentricities and inclinations of considered comets}
\begin{tabular}{lccc|cccc|cccc}
 & $a_\circ$ & $e_\circ$ &$i_\circ$ & & $a_\circ$ & $e_\circ$& $i_\circ$ & & $a_\circ$ & $e_\circ$& $i_\circ$\\
\hline
2P & 2.22 & 0.85 & $12^\circ$ & 9P & 3.12 &0.52&$10^\circ$ & 10P & 3.10 & 0.53 & $12^\circ$\\
 22P & 3.47&0.54&$4.7^\circ$ &28P & 6.91 & 0.78 & $14^\circ$ & 39P & 7.25&0.25&$1.9^\circ$ \\
44 P & 3.53 & 0.46& 7.0
\end{tabular} 
\end{table}

As the migration of TNOs to Jupiter's orbit was investigated
by several authors, we have made a series of simulations of the orbital evolution 
of JCOs under the gravitational influence of 
planets. We omitted the influence of Mercury 
(except for Comet 2P) and Pluto.  The orbital evolution of about 9352 and 
10301 JCOs with initial 
periods $P_a$$<$20 yr was integrated with the use of the Bulirsh-Stoer 
(BULSTO code [19]) and 
symplectic (RMVS3 code) methods, respectively.
We used the integration package of Levison and Duncan [20]. %(1994).  

      In the first series of runs (denoted as $n1$) we calculated the 
evolution of 3100 JCOs moving in initial orbits close to those of 20 
real comets with period $5$$<$$P_a$$<$9 yr,
and in the second series of runs (denoted as $n2$) we considered 7250 JCOs 
moving in initial orbits close to those of 10 real comets (with numbers 
77, 81, 82, 88, 90, 94, 96, 97, 110, 113) with period 5$<$$P_a$$<$15 yr.
In other series of runs, initial 
orbits were close to those of a single comet (2P, 9P, 10P, 22P, 28P, 
39P, and 44P). In order to compare the orbital evolution of comets and asteroids,
we also investigated the orbital evolution of asteroids initially moving 
in the 3:1 and 5:2 resonances with Jupiter. 
The number of objects in one run usually was $\le$250. 

In most JCO cases  the time $\tau$ when 
perihelion was passed was varied with a step $d\tau$$\le$1 day
(i.e., $\nu$ was varied with a step $<$$0.2^\circ$). Near the $\tau$ 
estimated from observations, we used smaller steps. In most JCO cases 
the range of initial values of $\tau$ was less than several tens of 
days. For asteroids, we varied initial values of $\nu$ and the 
longitude of the ascending node from 0 to 360$^\circ$.
The approximate values of initial orbital elements
($a$ in AU, $i$ in deg)
are presented in Table 2.
We initially integrated the orbits for  $T_S$$\ge$10 Myr. After 10 
Myr we tested whether some of remaining objects could reach inside 
Jupiter's orbit; if so,  the
calculations were usually continued. Therefore the results for orbits 
crossing or inside Jupiter's orbit were the same as if the 
integrations had been carried to the entire 
lifetimes of the objects. For Comet 2P and 
resonant asteroids, we integrated  until all objects were ejected 
into hyperbolic orbits or collided with the Sun. In some previous 
publications we have used smaller $T_S$, so these new data are more 
accurate.

      In our runs, planets were considered as material points so 
literal collisions did not occur.  However, using the formulas of the 
previous section, and the orbital elements sampled with a 500 yr 
step, we calculated the mean probability  $P$ of collisions. We 
define $P$ as  $P_\Sigma/N$, where $P_\Sigma$ is the probability for 
all $N$ objects of a
collision of an object with a planet during its lifetime,   the mean 
time $T$=$T_\Sigma/N$ during which
perihelion distance $q$ of an object was less than the semi-major 
axis $a_{pl}$ of the planet,
the mean time $T_d$ (in Kyr) spent in orbits with $Q$$<$4.2 AU,
and the mean time $T_J$ during which an object moved in 
Jupiter-crossing orbits. The values of $P_r$=$10^6P$,
$T_J$, $T_d$, and $T$ are shown in Tables 3-4. Here $r$ is the ratio of the 
total time interval when orbits are of Apollo
type ($a$$>$1 AU, $q$=$a(1-e)$$<$$1.017$ AU) at $e$$<$0.999 to that 
of Amor type ($1.017$$<$$q$$<$1.3 AU) and $T_c$=$T/P$ (in Gyr).
In almost all runs $T$ was equal to the mean time in planet-crossing 
orbits and 1/$T_c$ was a probability
of a collision per year (similar to 1/$T_f$).

\begin{table*}
%\begin{center}
\begin{minipage}{16cm}

\caption[]{
Mean probability $P$$=$$10^{-6}P_r$ of a collision of an object with 
a planet (Venus=V, Earth=E, Mars=M)
during its lifetime,  mean time $T$ (in Kyr) during which 
$q$$<$$a_{pl}$, $T_c$=$T/P$ (in Gyr),
mean time $T_J$ (in Kyr) spent in Jupiter-crossing orbits, 
mean time $T_d$ (in Kyr) spent in orbits with $Q$$<$4.2 AU,
and ratio $r$ of times spent in Apollo and Amor orbits. 
Results from  BULSTO 
code at $10^{-9}$$\le$$\varepsilon$$\le$$10^{-8}$
(marked as $10^{-9}$) and 
at  $\varepsilon$$\le$$10^{-12}$ (marked as $10^{-12}$)
and with RMVS3 code at integration step $d_s$.
The series of runs with a few excluded objects 
that had the largest probabilities of collision with the Earth
are marked by $^*$.
% $\Sigma$ denotes the sum for
%several series presented in the above lines at $10^{-9}$$\le$$\varepsilon$$\le$$10^{-8}$. 
%For $N$=7349, 2P runs were excluded.
}

$ \begin{array}{lll|cccccccccl}
\hline

   & & &$V$ & $V$ &  $E$ & $E$ & $E$ & $M$ & $M$ &  & &\\

\cline{4-13}

 &\varepsilon $ or $ d_s & N& P_r & T & P_r & T &T_c& P_r & T & r & T_J &T_d \\ % [N_d]\\

\hline

n1&10^{-9}&1900& 2.42 & 4.23 & 4.51 & 7.94 &1.76&  6.15 & 30.0 & 0.7 & 119& 20 \\ %[43/7]\\
n1&\le$$10^d&1200&25.4 & 13.8 & 40.1 & 24.0 &0.60&2.48 & 35.2 
& 3.0 &117 & 25.7 \\ % [41/3]\\
n1&\le$$10^d&1199^*&7.88 & 9.70 & 4.76 & 12.6 &2.65&0.76 & 16.8 
& 2.8 &117 &10.3 \\ % [40/2]\\
\hline
n2&10^{-9}&1000&10.2&27.5&14.7&43.4&2.95&2.58&62.6&3.1&187&8.3\\
n2&\le$$10^d&6250&17.9&28.2&17.3&40.8&2.36&3.17&63.1&3.2&147&26.5\\
\hline
$2P$&10^{-9} & 501^* & 141 &345 & 110 & 397 &3.61&10.5& 430 & 18.& 173&249 \\

$2P$&10^{-12}&100 &321&541&146&609&4.2   &14.8 & 634 & 27. &20 & 247\\ % [100/100]\\
$2P$&10^d & 251 &860&570&  2800 &788 & 0.28 &294 & 825 & 22. &0.29&614 \\ %[251/251]\\
$2P$&10^d & 250^* &160&297&  94.2 & 313 & 3.32 & 10.0&324 &35. &0.29& 585\\ % [250/250]\\
\hline
$9P$&10^{-9} & 800&1.34 &1.76 &3.72 & 4.11 &1.10&0.71 & 9.73& 1.2 &96& 2.6 \\ %[13/3]\\
$9P$&10^d & 400 &1.37&3.46&3.26&7.84&2.40&1.62&23.8& 1.1 &128& 8.0 \\ %[13/3]\\
\hline
$10P$&10^{-9} & 2149^*&28.3 & 41.3& 35.6 & 71.0&1.99&10.3 & 169.&1.6 &122&107\\ % [412-112]\\

$10P$&\le$$10^d &450&14.9&30.4&22.4&41.3&1.84&6.42& 113.&1.5 &85&44.\\ % [70/24]\\
\hline
$22P$& 10^{-9} &1000&1.44&2.98&1.76&4.87&2.77&0.74 & 11.0& 1.6 &116&1.5 \\ %[10/1]\\
$22P$&10^d & 250& 0.68 &2.87&1.39&4.96&3.57&0.60 & 11.5& 1.5 &121 & 0.6 \\ %[3/0]\\
\hline
$28P$&10^{-9} & 750 & 1.7 & 21.8& 1.9 & 34.7&18.3& 
0.44 & 68.9&1.9 &443&0.1 \\ %[8/0]\\
$28P$&10^d & 250 & 3.87 & 35.3& 3.99 & 59.0 &14.8& 0.71 & 109. 
&2.2 &535 &3.3 \\ %[7/0]\\
\hline
$39P$&10^{-9} & 750 & 1.06&1.72&1.19& 3.03&2.55& 
0.31 & 6.82& 1.6 &94&2.7 \\ %[10/1]\\
$39P$&10^d & 250 & 2.30 &2.68&2.50& 4.22 & 1.69 & 0.45 & 7.34& 
 2.2 &92 & 0.5 \\ %[2/0]\\
\hline
$44P$&10^{-9} & 500 &2.58&15.8&4.01&24.9&6.21&0.75&46.3&2.0&149&8.6\\
$44P$&10^d&1000&3.91&5.88&5.84&9.69&1.66&0.77&16.8&2.3&121&2.9\\
\hline
%\Sigma & 10^{-9} &7349& 9.52 & 16.2& 12.6 & 27.9&2.21 &  4.89 & 
%62.4& 1.4 &112&41 \\
%\Sigma &10^{-9} & 7850& 17.9 & 37.7& 18.8 & 51.5&2.74 &  5.29 & 
%85.9& 2.6 &116&54 \\
%\Sigma &10^{-9} & 7852& 130. & 72.6 & 84.5 & 95.7 & 1.13 & 13.5 & 
%132& 3.8 &116&101 \\
%%$R2$&10^{-9} & 24 &0.53&0.6&2.84&1.6&0.56&6.97&14&2.0&&\\
%\hline

\end{array} $

\end{minipage}

\end{table*}

\begin{table*}
%\begin{center}
\begin{minipage}{16cm}

\caption[]{  Same as for Table 2, but for resonant asteroids.
$i_o$=10$^o$ at $e_o$=0.15, and $i_o$=5$^o$ at $e_o$=0.05.
}

$ \begin{array}{llll|cccccccccl}
\hline

  && & &$V$ & $V$ &  $E$ & $E$ & $E$ & $M$ & $M$ &  & &\\

\cline{5-14}

 &e_o&\varepsilon $ or $ d_s & N& P_r & T & P_r & T &T_c& P_r & T & r & T_J &T_d \\ % [N_d]\\

\hline

3:1&0.15&10^{-9}& 288  &1286 &1886& 1889& 2747 &1.45& 
488&4173 & 2.7 &229&5167 \\
3:1&0.15&\le$$10^{-12}& 70 & 1162 &1943& 1511& 5901 &3.91& 587& 
803& 4.6 &326 & 8400 \\% [70/70]\\
3:1&0.15&10^d& 142^* &27700&8617& 2725& 9177&3.37& 1136&9939& 16. &1244&5000 \\% [142/140]\\
\hline
5:2&0.15&10^{-9}& 288 &101   &173& 318 & 371 &1.16&209 
&1455 & 0.5 &233& 1634 \\
5:2&0.15&10^{-12}& 50 & 130   &113& 168 & 230 &1.37&46.2 &507 
& 1.4 &166 &512 \\%[50/4]\\
5:2&0.15&10^d& 144 & 58.6&86.8&86.7&174 &2.01&17&355&1.7 &224& 828\\% [144/13]\\

\hline
3:1&0.05&10^{-9}& 144 & 200 &420&417 &759  &1.82& 195&1423&2.1 &157&2620 \\
3:1&0.05&10^d& 144 & 10051&2382&6164&4198&0.68&435&5954&2.5&235&18047\\
5:2&0.05&10^{-9}& 144 & 105 &114& 146& 214 &1.47& 42&501& 1.5 &193&996 \\
5:2&0.05&10^d& 144 &148&494&173&712&4.12&51&1195&2.3&446&984\\
\hline

\end{array} $

\end{minipage}

\end{table*}

In Table 5 for several objects that had large collision probabilities
with the Earth, we present times (Myr) spent by these objects 
in orbits typical for
inner-Earth objects (IEOs, $Q$$<$0.983 AU), Aten ($a$$<$1 AU and
$Q$$>$0.983 AU), Al2 ($q$$<$1.017 AU and 1$<$$a$$<$2 AU), Apollo, and 
Amor objects, and also probabilities of collisions
 with Venus ($p_v$), Earth ($p_e$), and Mars ($p_m$)
during their lifetimes $T_{lt}$ (in Myr).
Objects 44P and 113P presented in Table 5 were not included 
in the lines $n1$ and $n2$ marked by $^*$ in Table 3, 
respectively,
and objects 2P and 10P with BULSTO were not included
in the lines for the corresponding series of runs in Table 3 with 
$N$=250 for 2P and $N$=2149 for 10P.

\section*{ORBITAL EVOLUTION OBTAINED BY DIRECT INTEGRATIONS }

%As the next step in estimating probabilities, we calculated the 
%orbits for thousands of test particles with orbits similar to known 
%comets and asteroids, but having slightly different initial 
%conditions.  The results confirm that most of the collision 
%probability comes from a handful of very rare cases in which the test 
%particle 
%is Earth-crossing for an 
%extended period of time.  They also show that the initial conditions 
%matter more than the choice of orbit integrator.

Here and in Figs. 1, 2a-c, 3-5 we present the results obtained by
the Bulirsh-Stoer method (BULSTO code [19]) with the integration step 
error less than
$\varepsilon$$\in$[$10^{-9}$-$10^{-8}$], and in the next 
section we compare them with those of BULSTO with 
$\varepsilon$$\le$$10^{-12}$ and  a symplectic method. 

\begin{table}[h]    % Table 5

\begin{center}
\begin{minipage}{15.7cm}
\caption{Times (Myr) spent by six objects in various 
orbits, and probabilities of collisions
 with Venus ($p_v$), Earth 
($p_e$), and Mars ($p_m$)
during their lifetimes $T_{lt}$ (in Myr)}
$ \begin{array}{lllll lllll l}
\hline

   & d_s $ or $ \varepsilon&$IEOs$& $Aten$&  $Al2$& $Apollo$ & $Amor$ 
& T_{lt} &p_v&p_e&p_m\\
\hline
%\cline{6-11}
$2P$&10^{-8}&0.1&83&249&251&15&352&0.224&0.172&0.065\\
$10P$&10^{-8}&10&3.45&0.06&0.06&0.05&13.6&0.665&0.344&0.001\\
$2P$ & 10^d & 12 & 33.6 & 73.4 & 75.6 & 4.7 & 126 &0.18&0.68&0.07\\
%$2P$ & 10^{-12} &  &  &  &  &  &  &&ADD??&11-20\\
$44P$ & 10^d&  0 & 0    & 11.7 & 14.2 & 4.2 & 19.5 &0.02&0.04&0.002\\
$113P$ &6^d&0 &0 & 56.8& 59.8&4.8& 67& 0.037& 0.016&0.0001\\
%3:1 & 10^d &  & & & & & 253 &&&ADD\\
$resonance $ 3:1 & 10^{-12} & 0 & 0 & 20 & 233.5 & 10.4 & 247 
&0.008&0.013&0.0007\\

\hline

\end{array}$
\end{minipage}
\end{center}
\end{table}

      The  results showed that most of the probability of collisions of
former JCOs with the terrestrial planets is due to a small 
($\sim$0.1-1\%) fraction that orbited for several Myr with aphelion 
$Q$$<$4.7 AU.
Some had typical asteroidal and NEO orbits and reached $Q$$<$3 AU for 
several Myr. 
Time variations in orbital elements of JCOs obtained by the BULSTO code are 
pre\-sen\-ted in Figs. 1, 2a-b. Plots in Fig. 1 are more typical than those in 
Fig. 2a-b, which were obtained  
for two JCOs with the highest probabilities of collisions with the terrestrial planets.
%for several JCOs and 
Fig. 2c shows the plots for an 
asteroid from the 3:1 resonance with Jupiter. 
The results obtained by a 
symplectic code for two JCOs are presented in Fig. 2d-e.
%The ratio of the mean probability of a collision of a JCO with $a$$>$1 AU
%with a planet to the mass of the planet was greater for Mars than
%that for Earth by a factor of several (
Large values of    
$P$ for Mars in the $n1$ runs were caused by a single object with a 
lifetime of 26 Myr. 

 The total times for Earth-crossing objects were mainly due to 
a few tens of objects    with high collision probabilities.
Of the JCOs with initial orbits close to those of 
10P and 2P, six and nine respectively moved into Apollo orbits with 
$a$$<$2 AU (Al2 orbits) for at least 0.5 Myr each, and five of them 
remained in such orbits for more than 5 Myr each. The contribution of 
all the other objects to Al2 orbits was smaller. 
Only one and two JCOs reached IEO and Aten orbits, respectively.

%%%
%
\begin{table}[h]   % Table 6
\begin{center}
\caption{
Times (in Myr) spent by $N$ JCOs and 
asteroids during their lifetimes, with results for first 50 Myr in  [ 
]. 
%Results from  BULSTO code at $10^{-9}$$\le$$\varepsilon$$\le$$10^{-8}$.
}
$ \begin{array}{lll|llllll}

\hline

   &$Method$& N& $IEOs$& $Aten$&  $Al2$& $Apollo$ & $Amor$\\

%\cline{6-11}
\hline
$JCOs$ &$BULSTO$& 9352 & 10 & 86 & 412 & 727 & 192 \\ %7100

$JCOs without 2P$&$BULSTO$& 8800 & 10 & 3.45 & 24 & 273& 165 \\ %7000
n1& $RMVS3$& 1200& 0& 0& 12 & 30 & 10\\
n2& $RMVS3$ &6250& 0& 0& 58&267&83\\
3:1 &$BULSTO$&  288   & 13  & 4.5 & 433 $ $[190] & 790 $ $ [540] & 290 $ $ 
[230] \\%& 83 $ $[78] \\

5:2 & $BULSTO$&288 & 0 & 0 & 17 $ $[2] & 113 $ $[90] & 211 $ $[90] \\%& 253 $ $[230]\\

\hline

\end{array} $
\end{center}

\end{table}

One former JCO (Fig. 2a), which had an initial orbit close to that of 10P, 
moved in Aten orbits for 3.45
Myr, and the probability of its collision with the Earth from such 
orbits  was 0.344 (so $T_c$=10 Myr was even smaller than the values of $T_f$
presented in Table 1; i.e., this object had smaller $e$ and $i$
than typical observed Atens),  greater
than that  for the 9350 other simulated former JCOs during 
their lifetimes (0.17). It also moved for
about 10 Myr  in inner-Earth orbits before its collision with Venus, 
and during this time the
probability $P_V$=0.655 of its collision with Venus was greater 
($P_V$$\approx$3 for the time interval presented in Fig. 2a) 
than that  for the 9350 JCOs during their lifetimes (0.15).
At $t$=0.12 Myr orbital elements of this object jumped considerably
and the Tisserand parameter increased from $J$$<$3 to $J$$>$6, and $J$$>$10
during most of its lifetime.
Another object (Fig. 2b) moved in highly eccentric Aten orbits for 83 Myr, and 
its lifetime before collision
with the Sun was 352 Myr. Its probability of collisions with Earth, 
Venus and Mars during its
lifetime was 0.172, 0.224, and 0.065, respectively. These two objects 
were not included in Table 3. % except for the entry for $N$=7852.
Ipatov [21] obtained the migration of JCOs into IEO and Aten 
orbits using the approximate
method of spheres of action for taking into account the gravitational 
interactions of bodies with planets.
%%The mean time $T_E$ during which a JCO was moving in Earth-crossing 
%%orbits is $9.6\times10^4$ yr for
%%the 7852 simulated JCOs, and $\approx8\times10^3$ yr for the $n1$ case.
%%The data for Comet P/1996 R2 (line R2) were not included in the sums in Table 2.
In the present paper we consider only the integration into the future.
Ipatov and Hahn [9] integrated the evolution of Comet P/1996 R2 both
into the future and into the past, in this case 
the mean time $T_E$ during which a JCO was moving in Earth-crossing orbits is
$T_E=5\times10^3$ yr.
%Earlier several scientists obtained smaller values of $P$ than those 
%presented in Table 2.
The ratio $P_S$ of the number of objects colliding with the Sun to 
the total number of
escaped (collided or ejected) objects was less than 0.015 
for the considered runs (except for 2P).

%+  return this text for UMD paper
% Need to place this table carefully to avoid being broken up
%\begin{table}[h]
  \begin{center}
%\begin{minipage}{17cm}
%%% \caption
{Ratio $P_S$ of objects colliding with the Sun 
  to those colliding with planets or ejected}
  $ \begin{array}{llllllll}
  \hline
  $Series$&n1 & 9P & 10P & 22P & 28P & 39P &44P\\
  P_S & 0.0005 & 0 & 0.014 & 0.002 & 0.007 & 0 &0.004\\
  \hline
  \end{array}$
%\end{minipage}
  \end{center}
%\end{table}

Some former JCOs spent a long time in the 3:1 resonance with Jupiter
(Fig. 1a-b)
and with 2$<$$a$$<$2.6 AU. Other objects reached Mars-crossing orbits 
for long times. We conclude that JCOs can supply bodies to the 
regions which are considered by many scientists [3] %(Bottke et al., 2002) 
to belong to the main sources of NEOs, and that those rare objects 
that make transitions to typical NEO orbits
dominate the statistics.  Only computations 
with very large numbers of objects can hope to reach accurate 
conclusions on collision probabilities 
with the terrestrial planets.

\begin{figure}
\includegraphics[width=81mm]{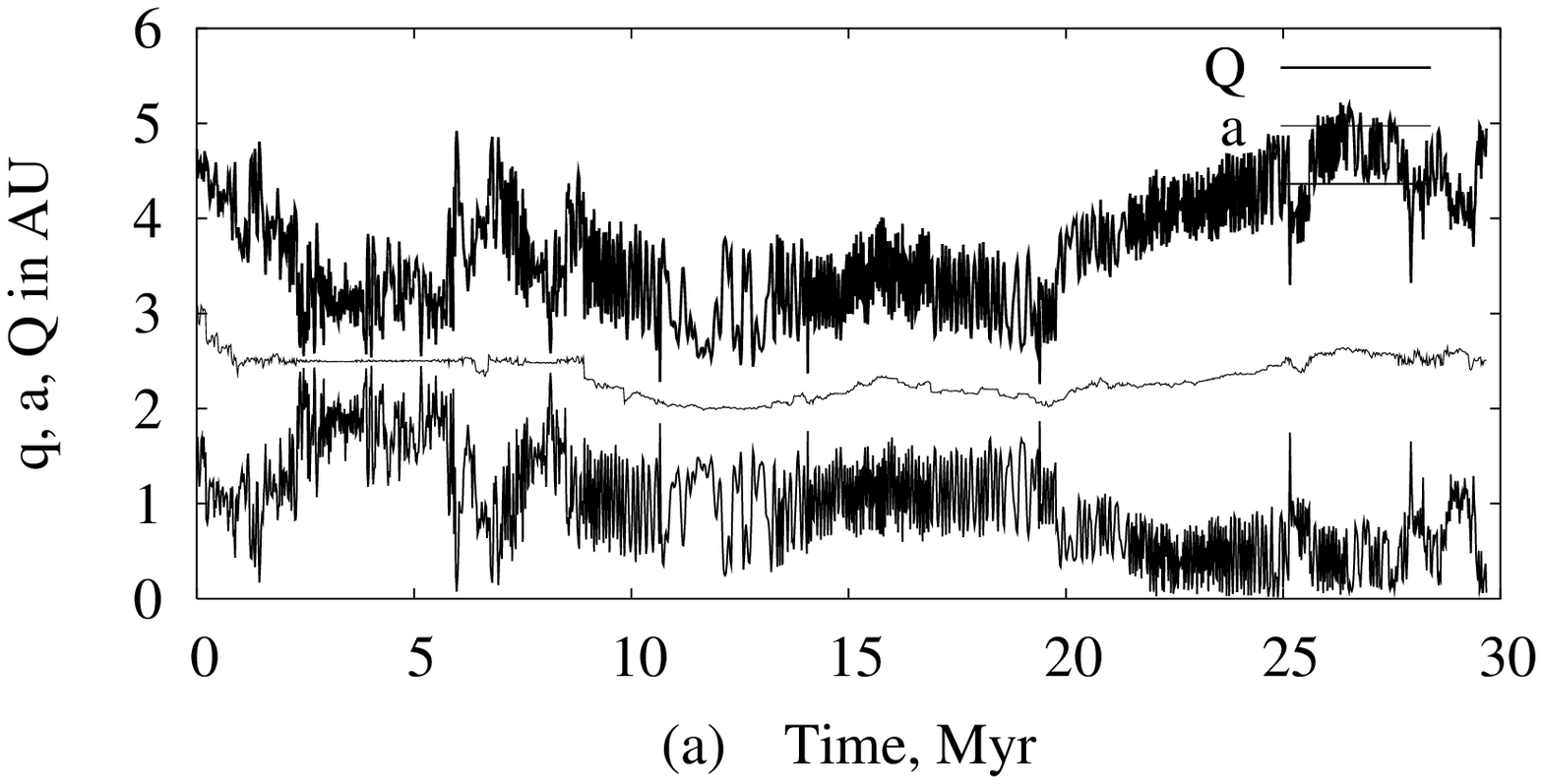}
\includegraphics[width=81mm]{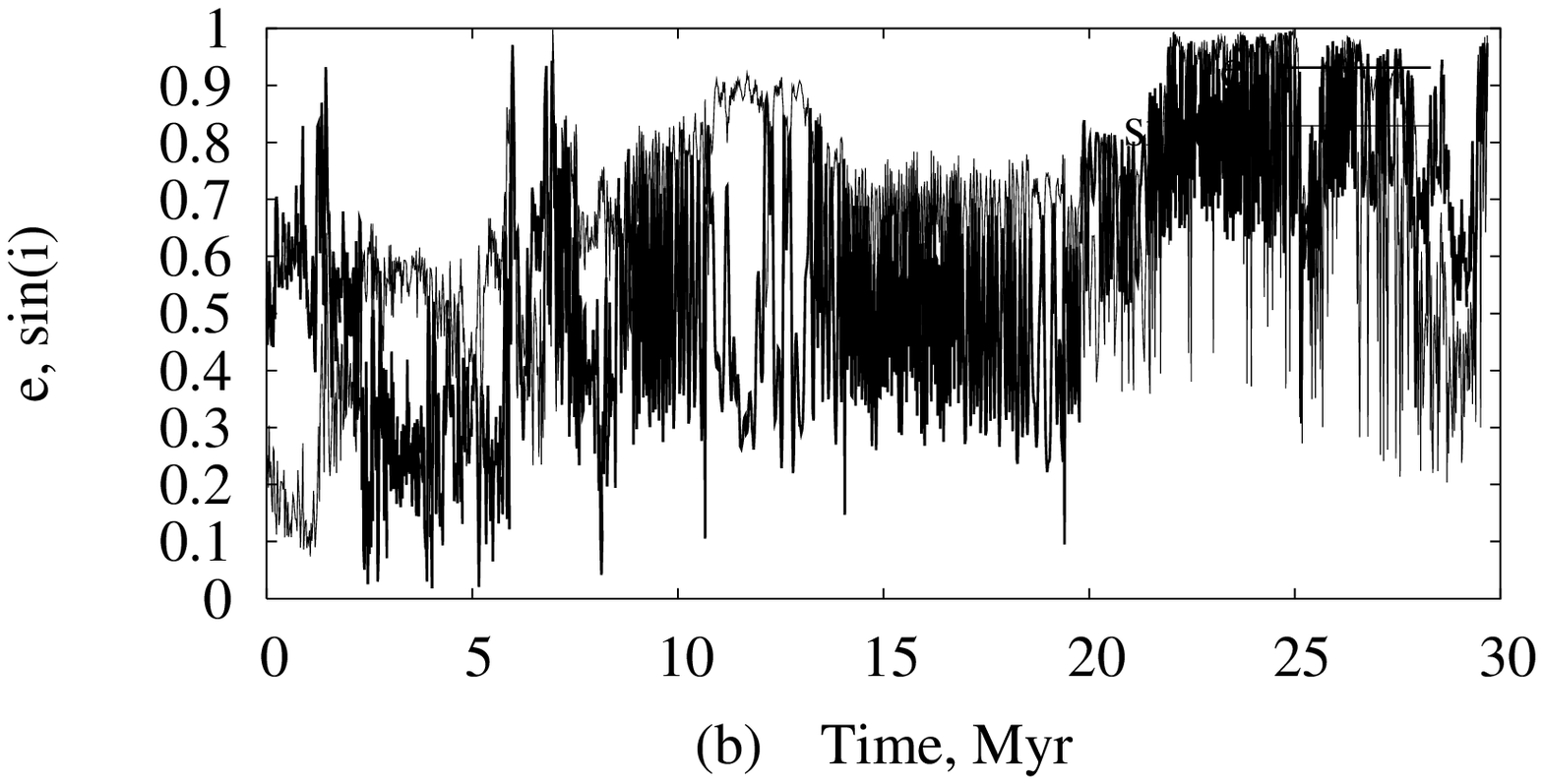}
\includegraphics[width=81mm]{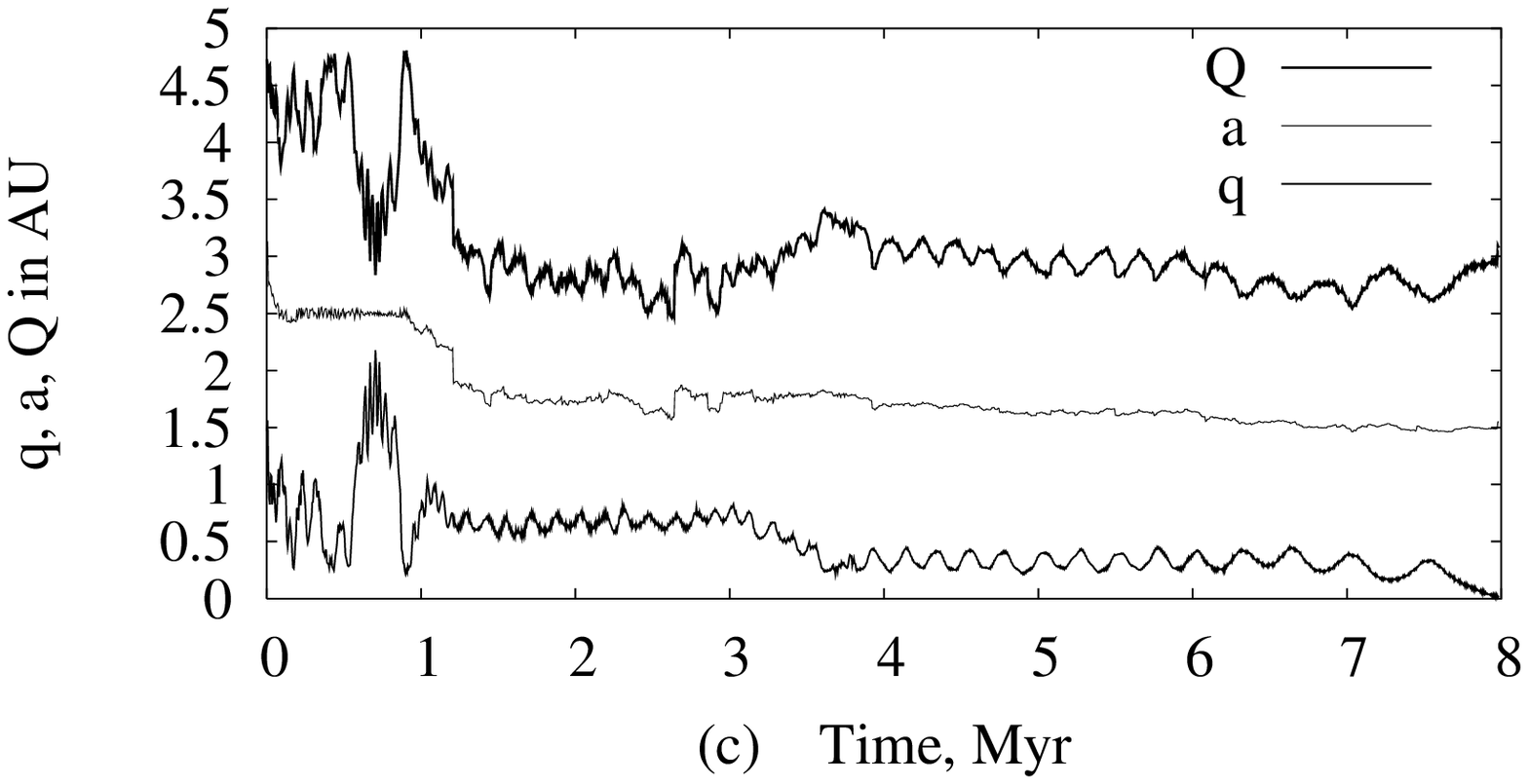}
\includegraphics[width=81mm]{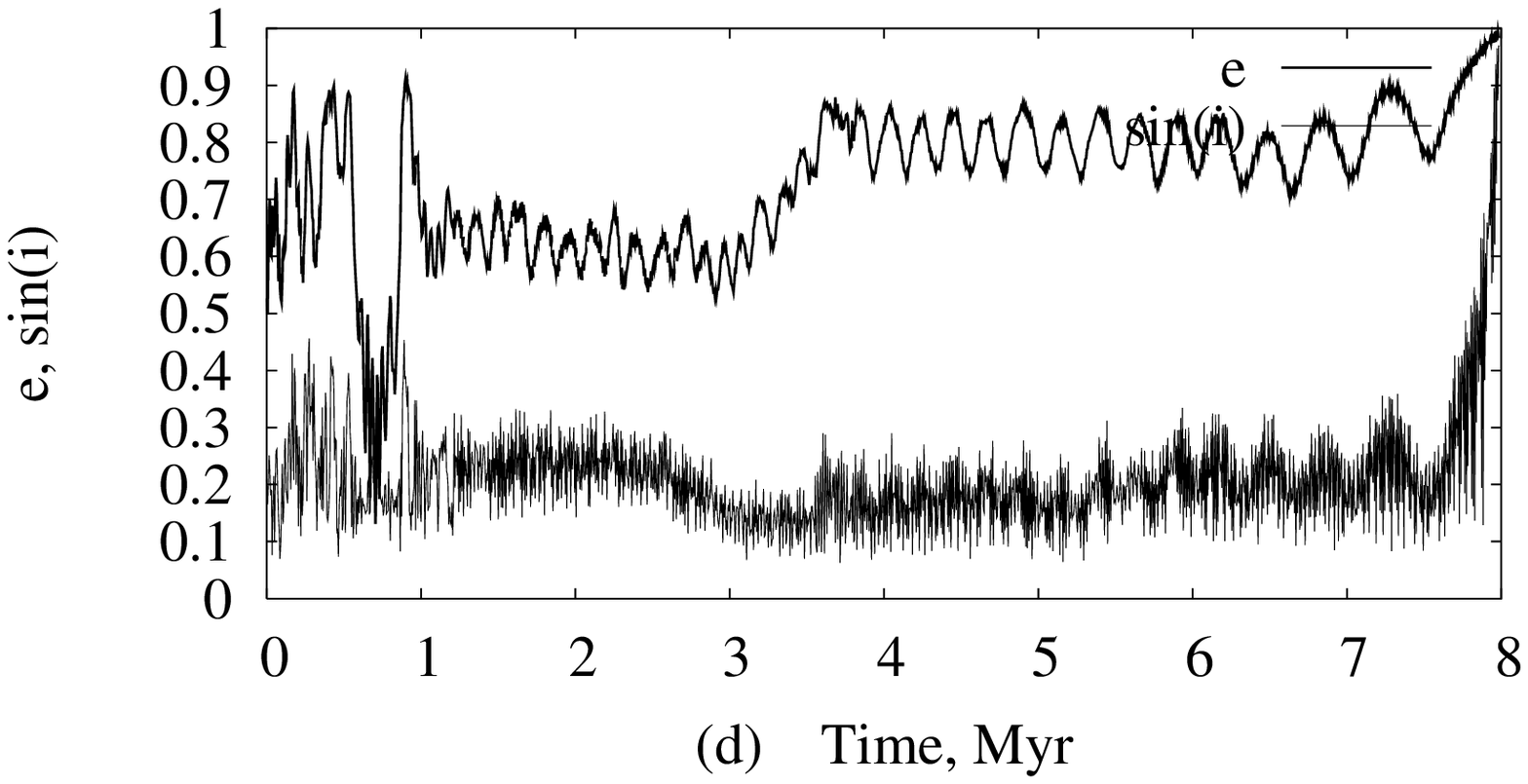}
\includegraphics[width=81mm]{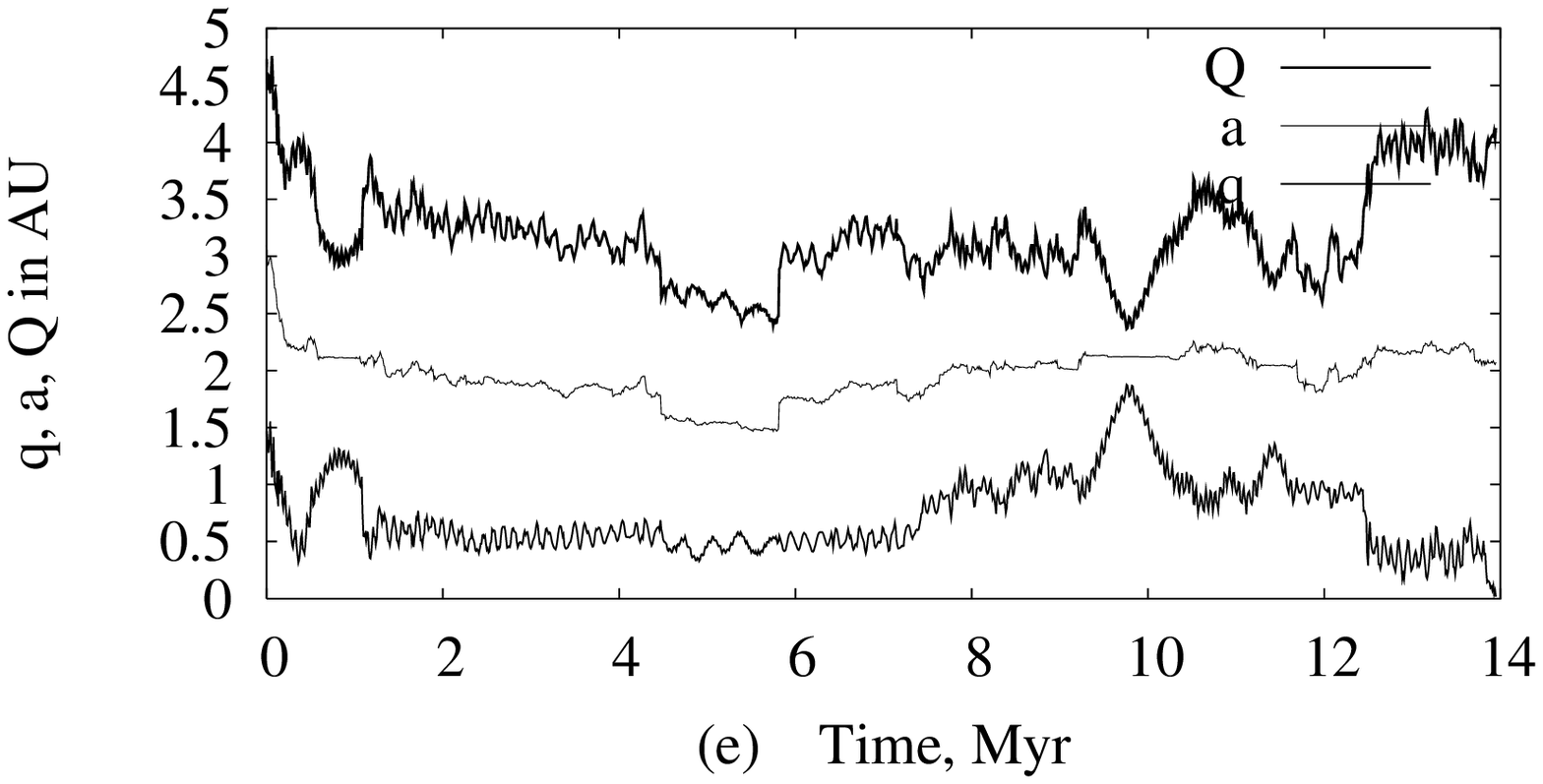}
\includegraphics[width=81mm]{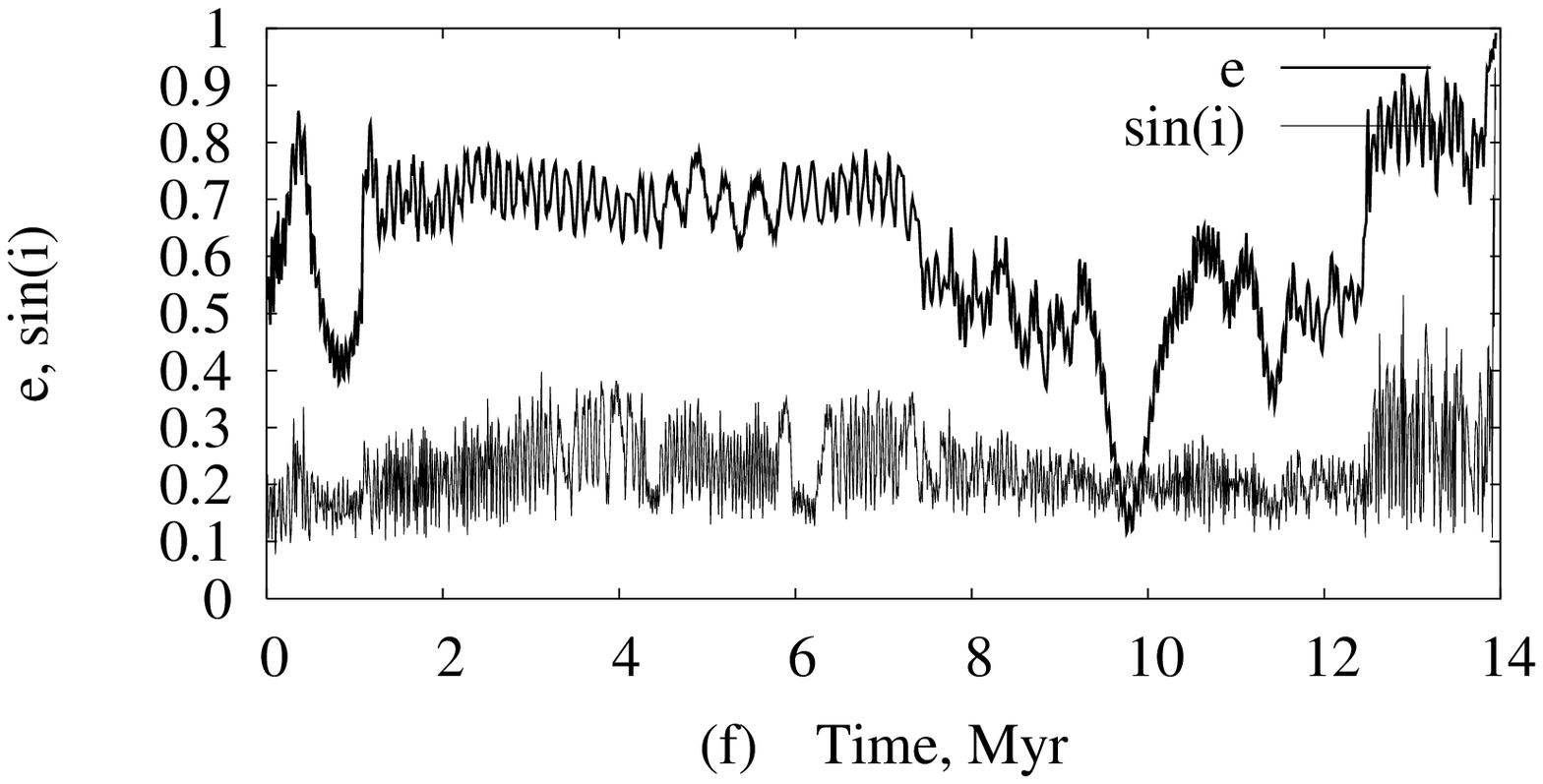}
\includegraphics[width=81mm]{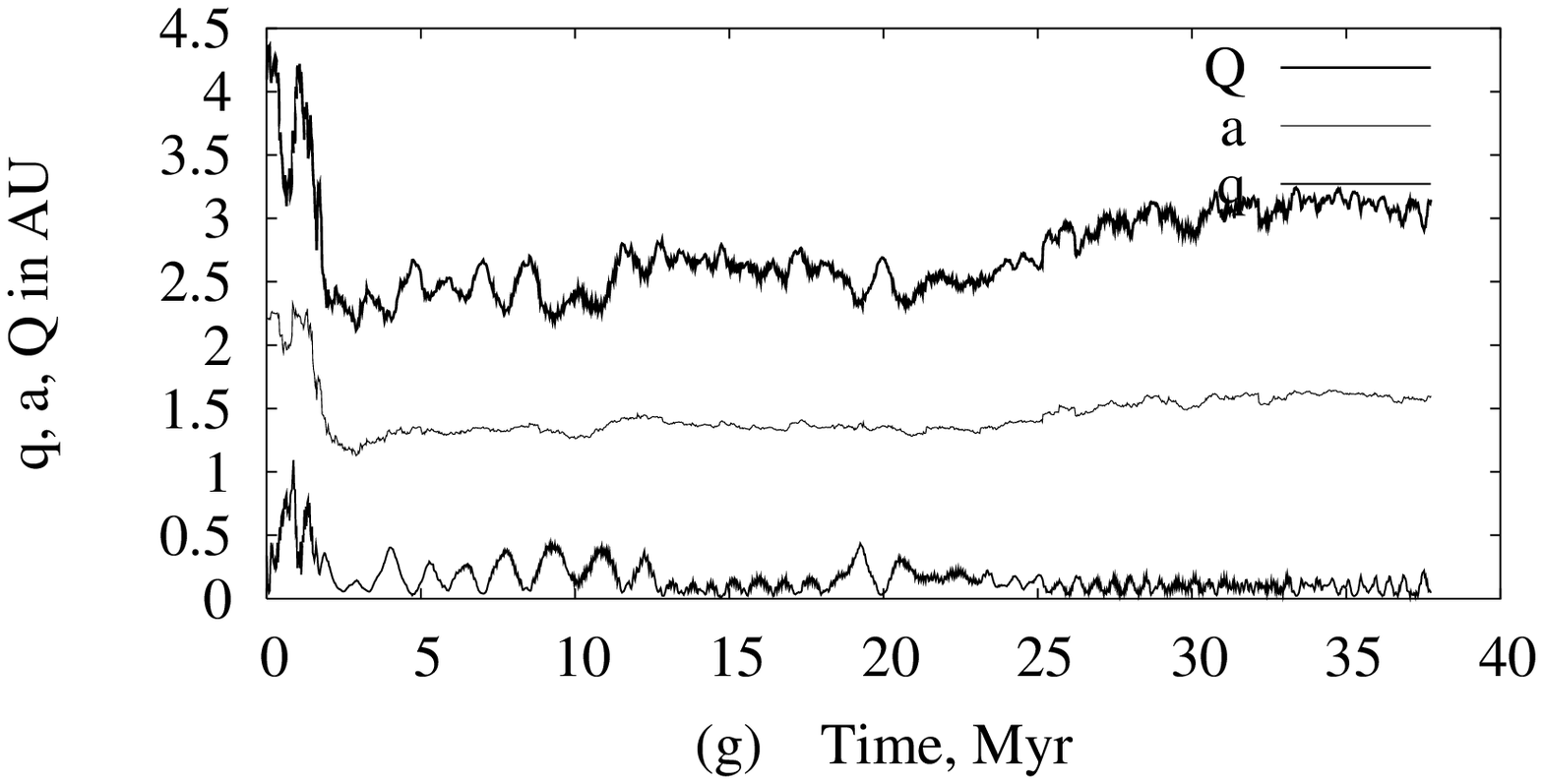}
\includegraphics[width=81mm]{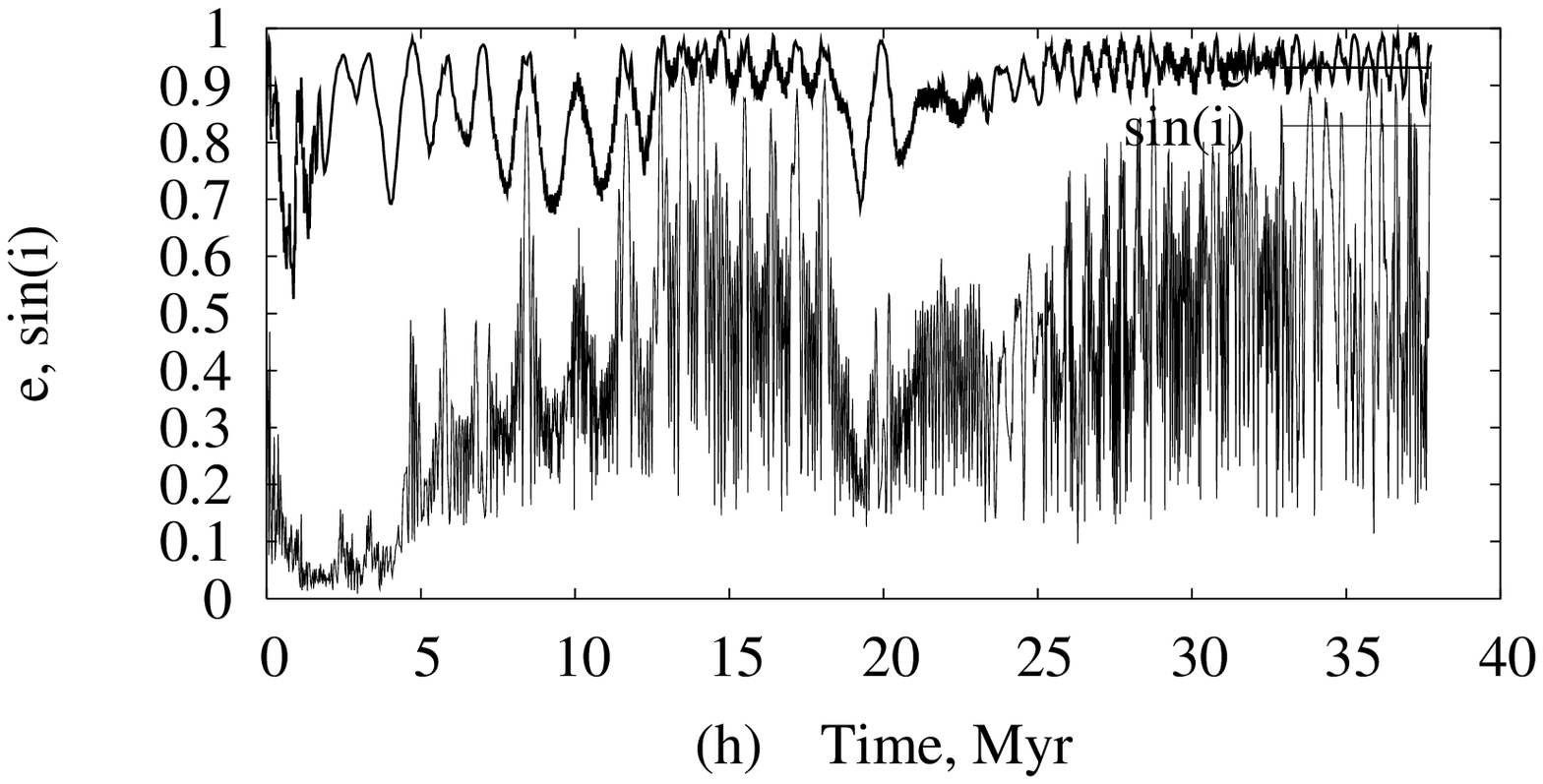}
\caption{Time variations in $a$, $e$, $q$, $Q$, sin($i$) for a former JCO in 
initial orbit close to that of Comet 10P (a-f), or Comet 2P (g-h). 
Results from BULSTO code with $\varepsilon \sim 10^{-9}-10^{-8}$.} 
\end{figure}%

\begin{figure}

\includegraphics[width=52mm]{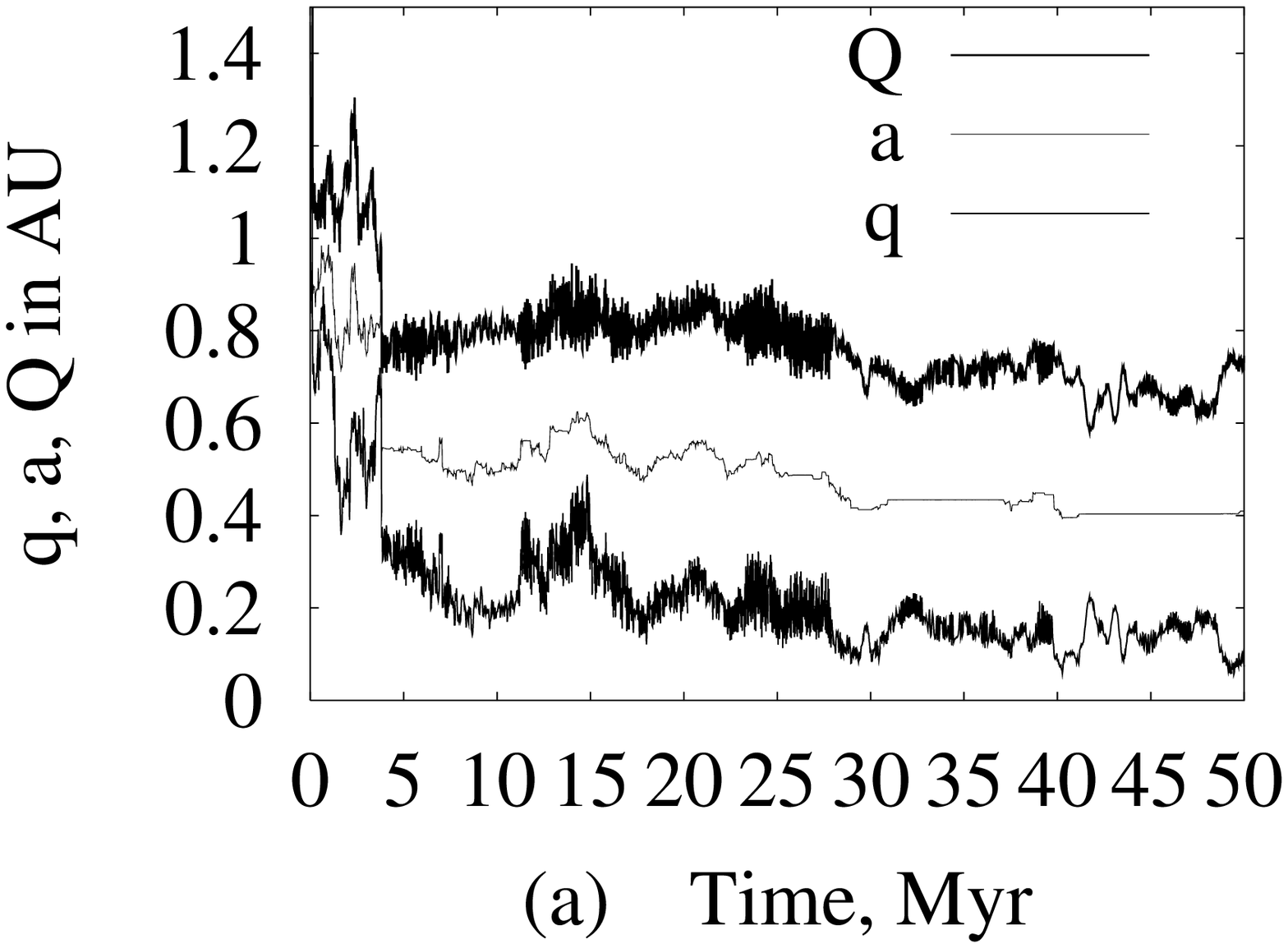}
\includegraphics[width=52mm]{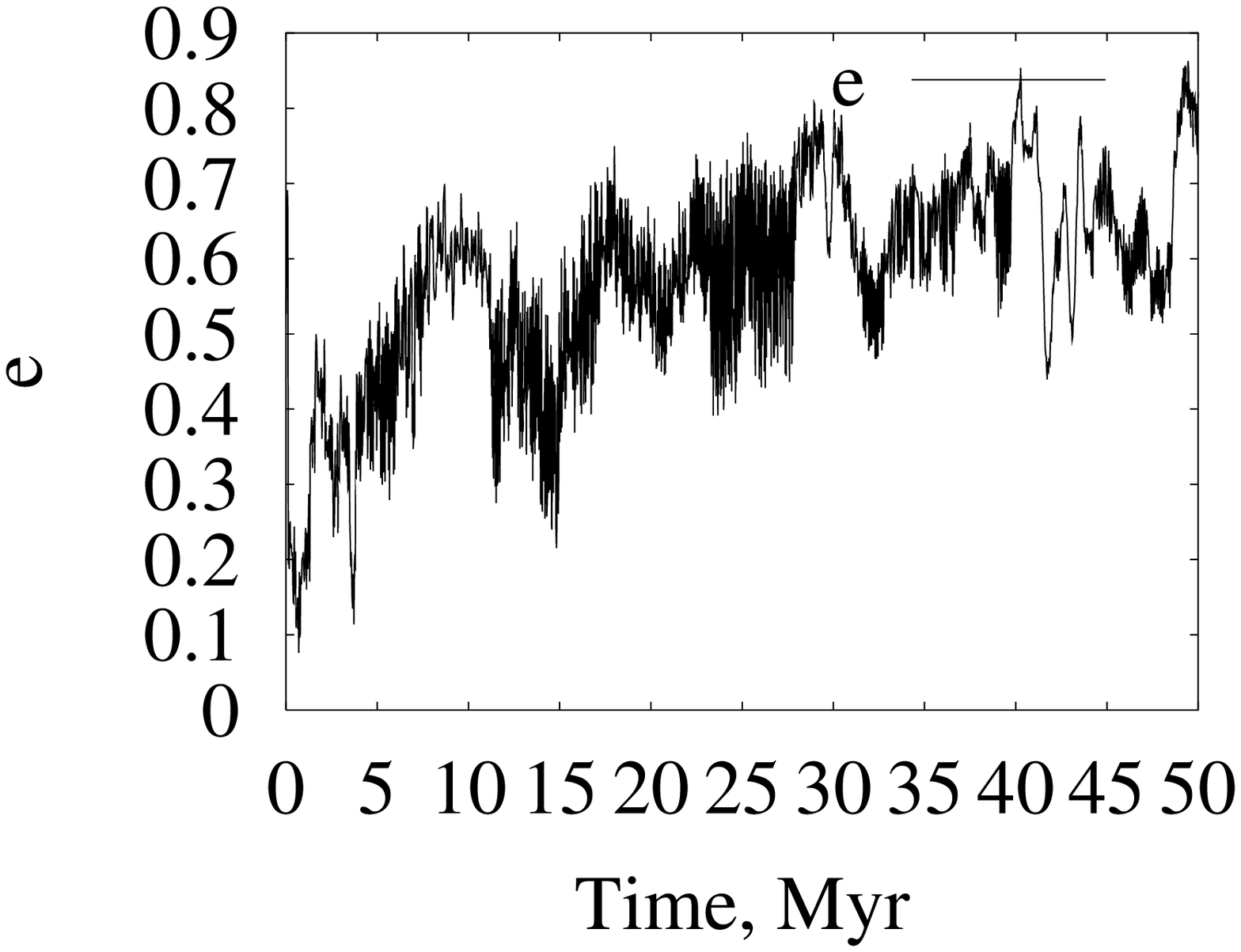}
\includegraphics[width=52mm]{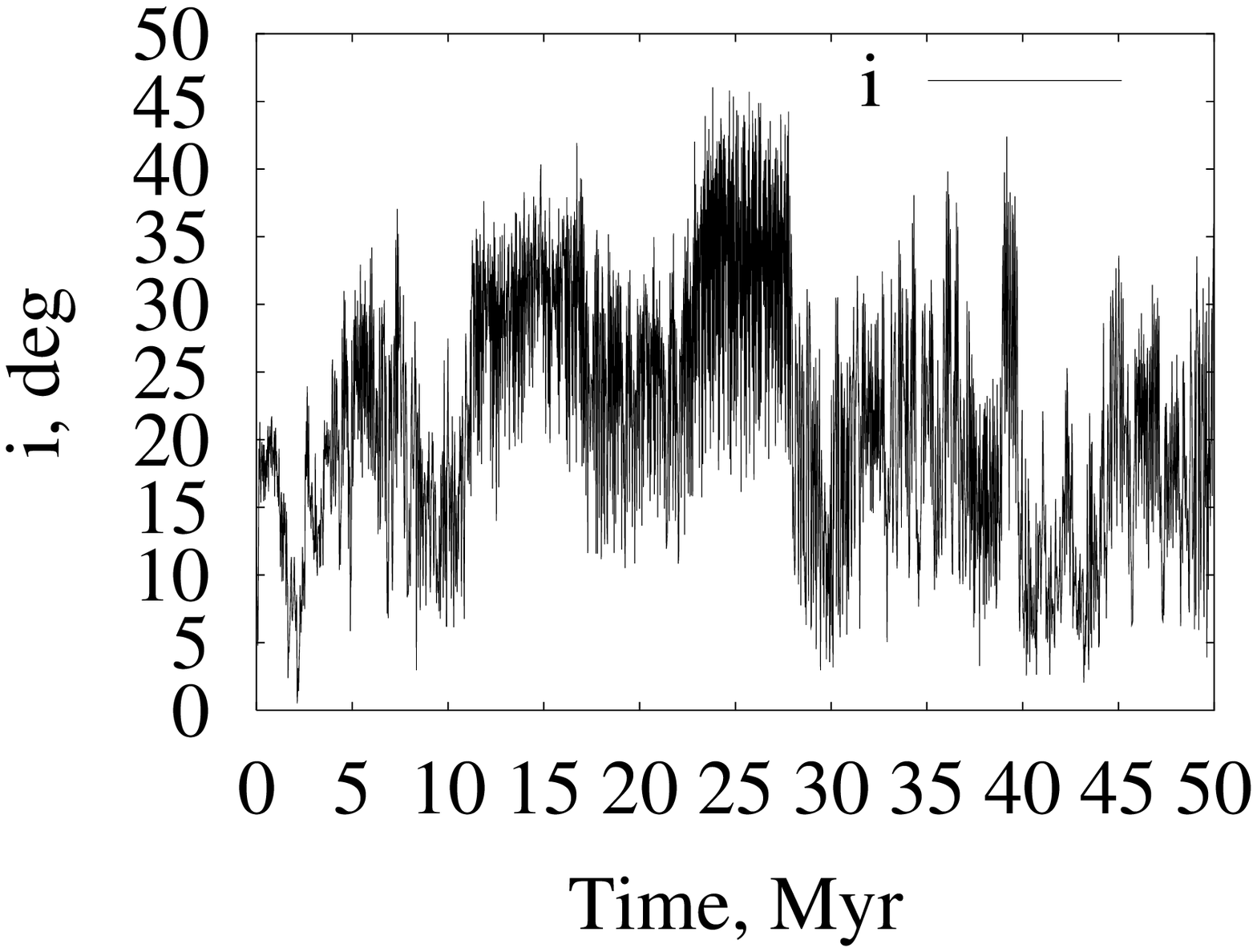}

\includegraphics[width=52mm]{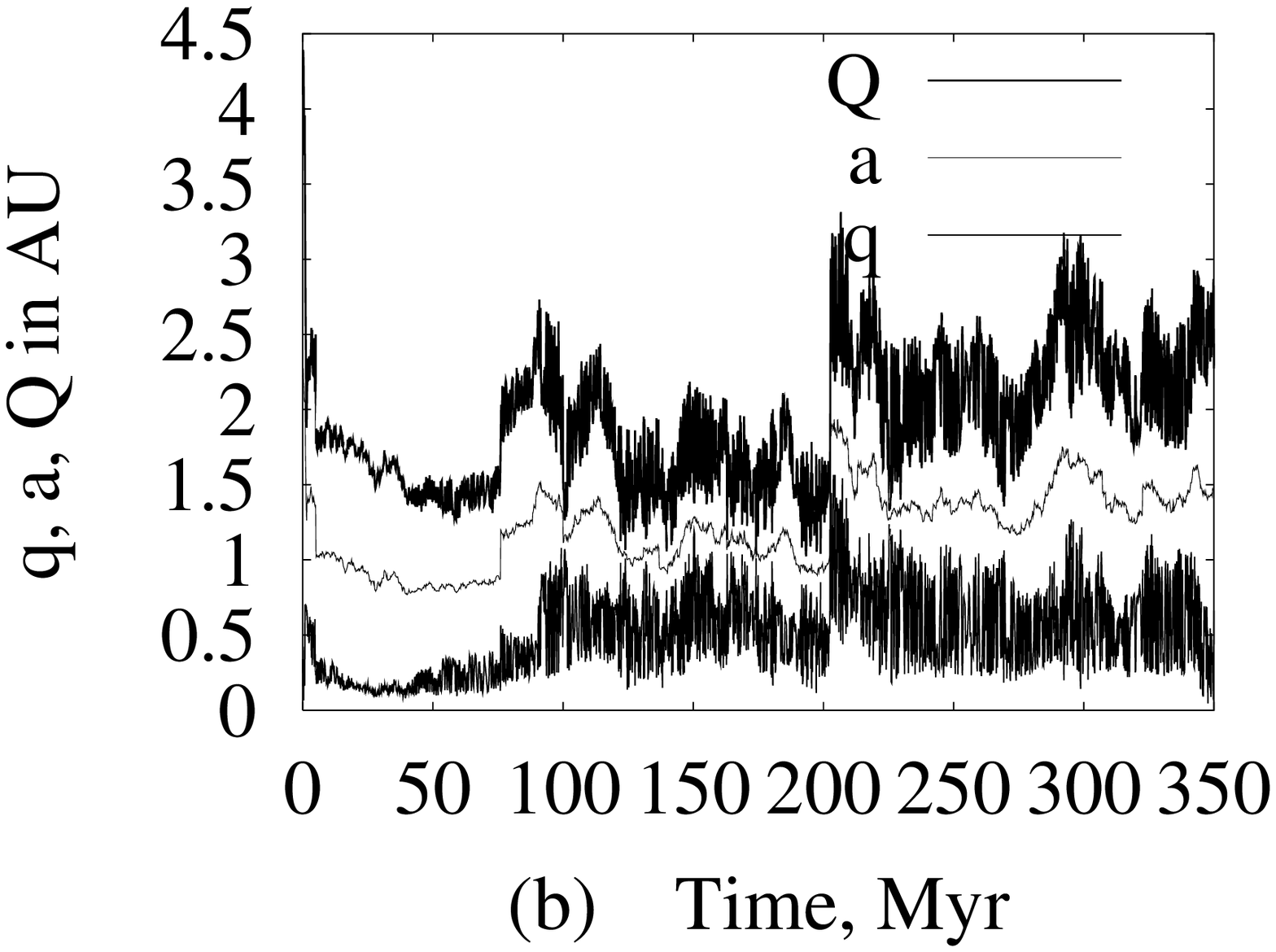}
\includegraphics[width=52mm]{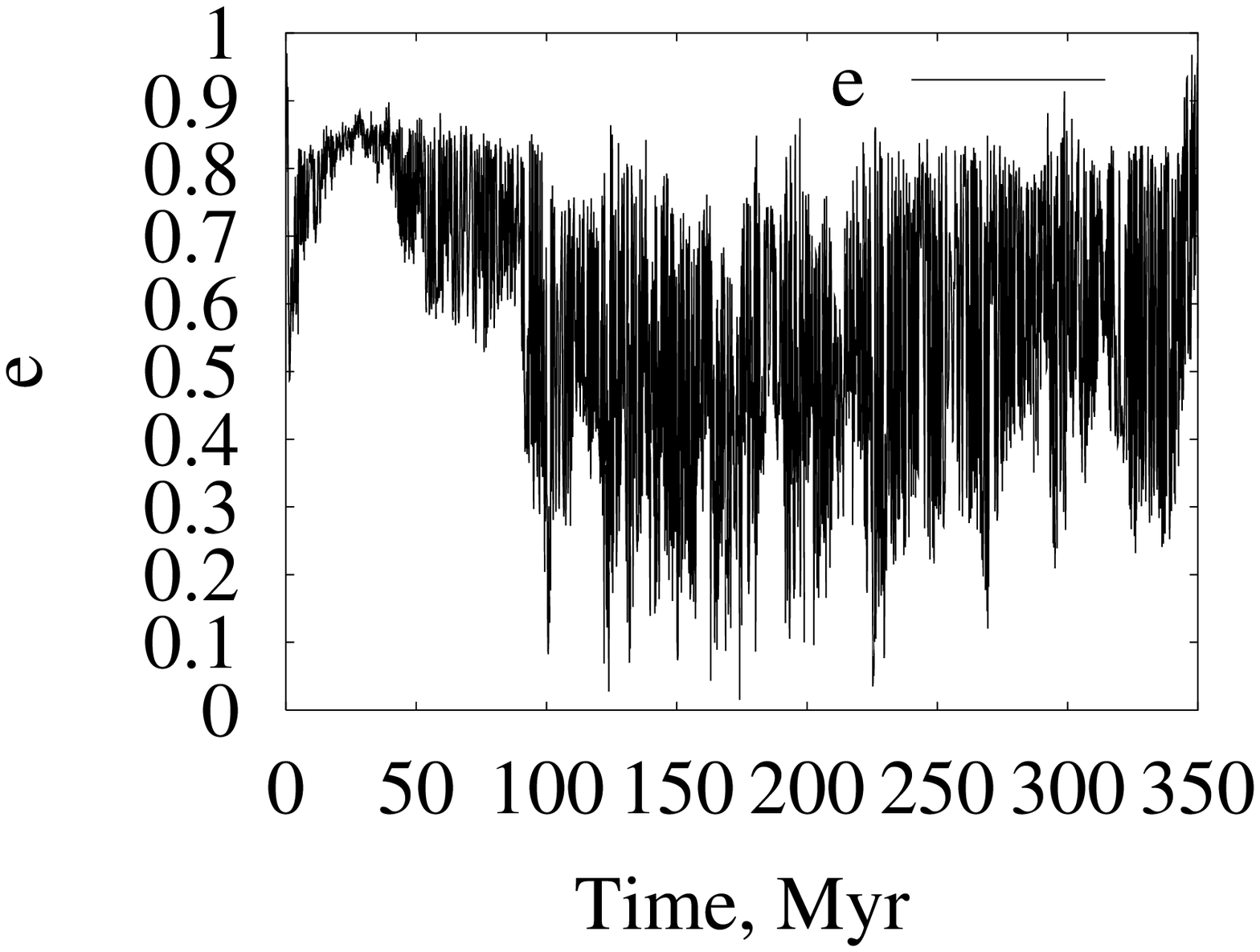}
\includegraphics[width=52mm]{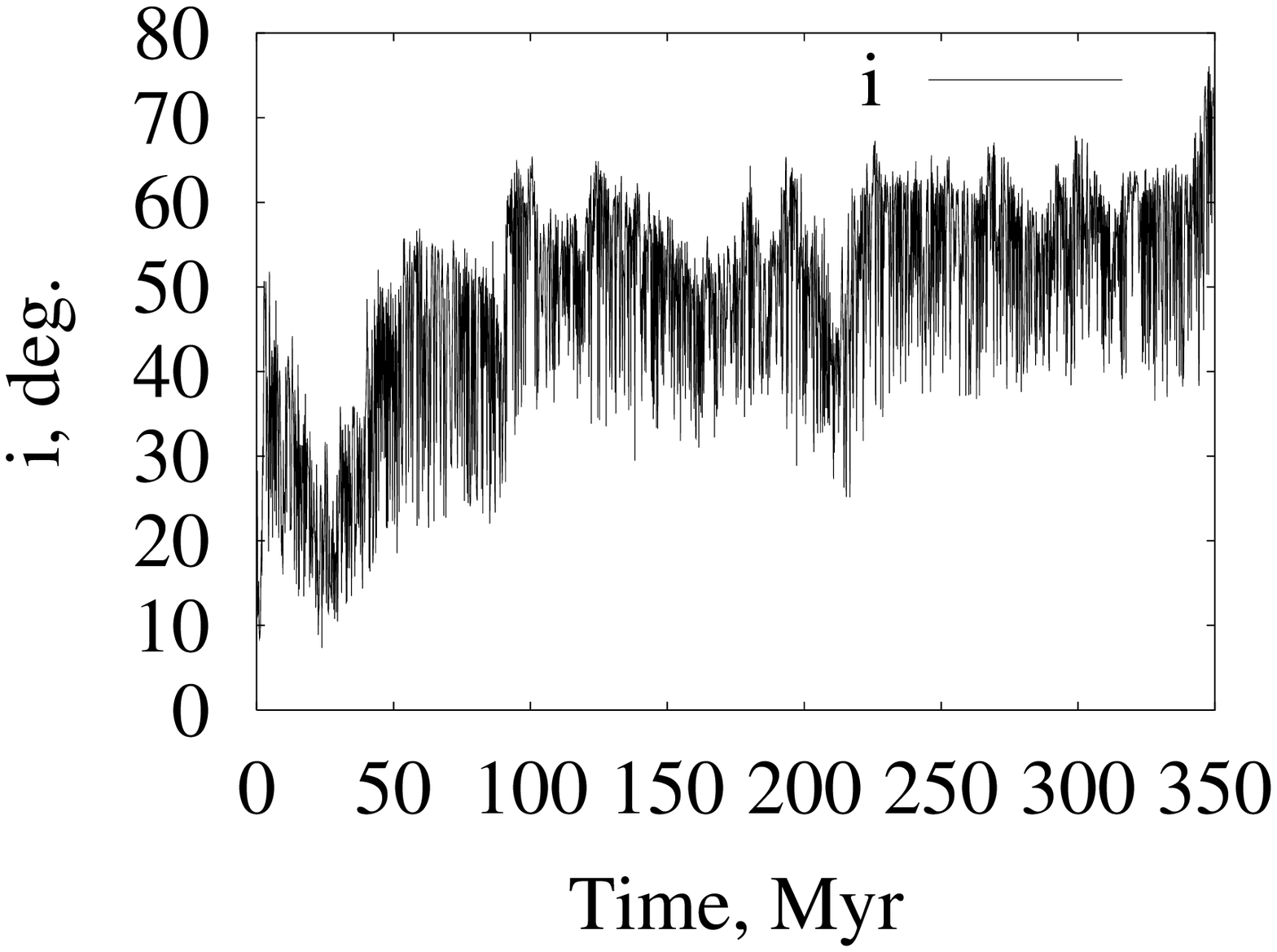}

\includegraphics[width=52mm]{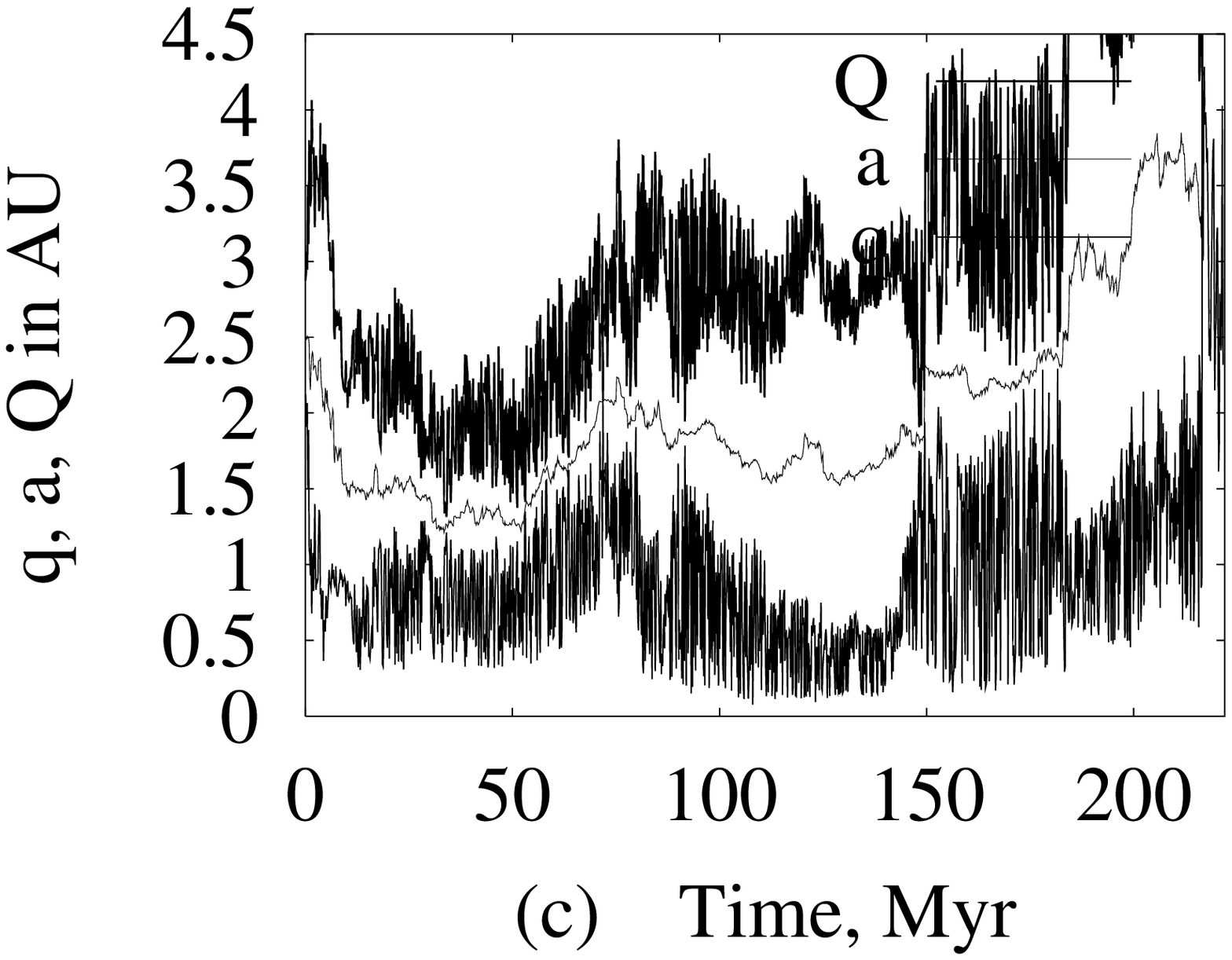}
\includegraphics[width=52mm]{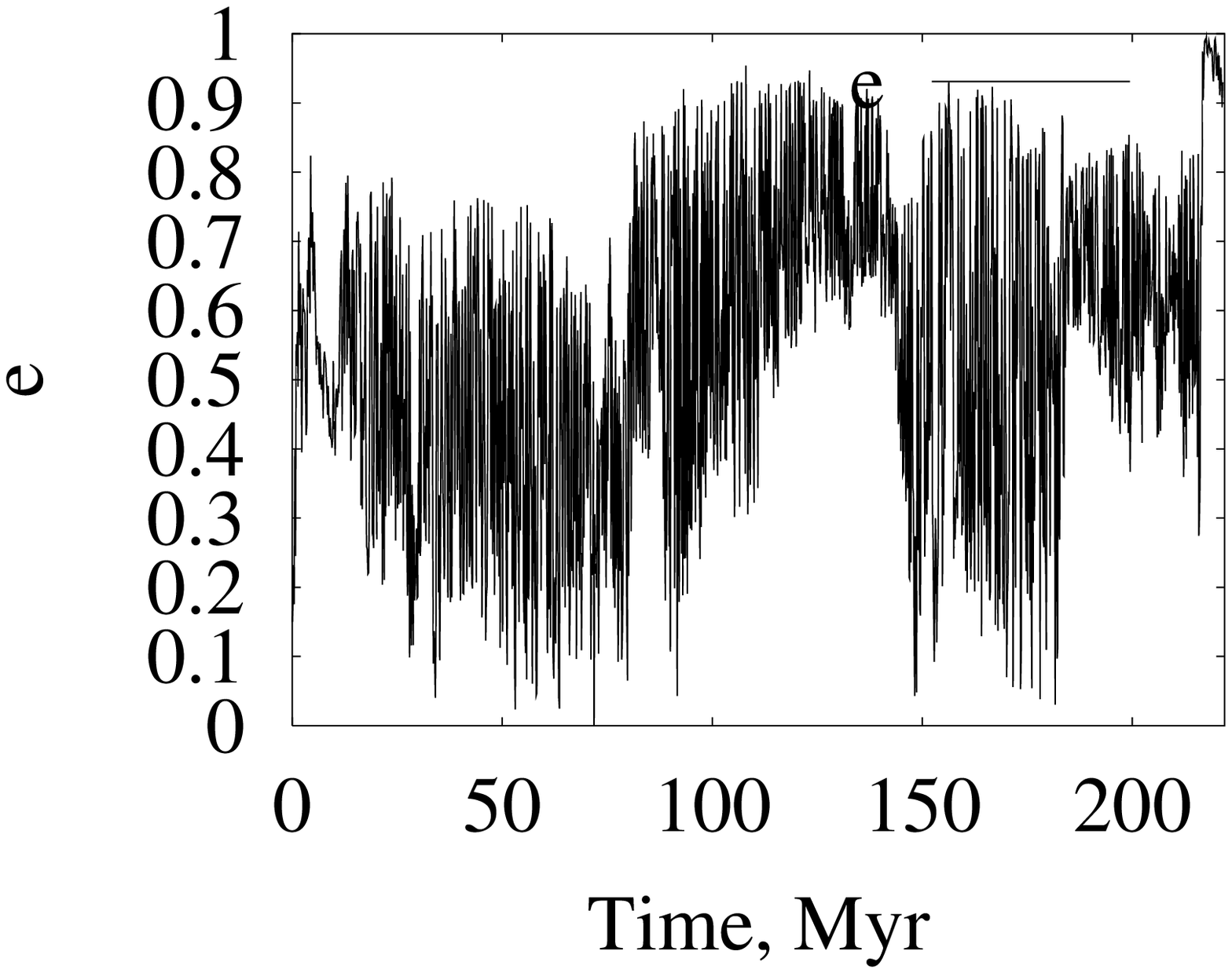}
\includegraphics[width=52mm]{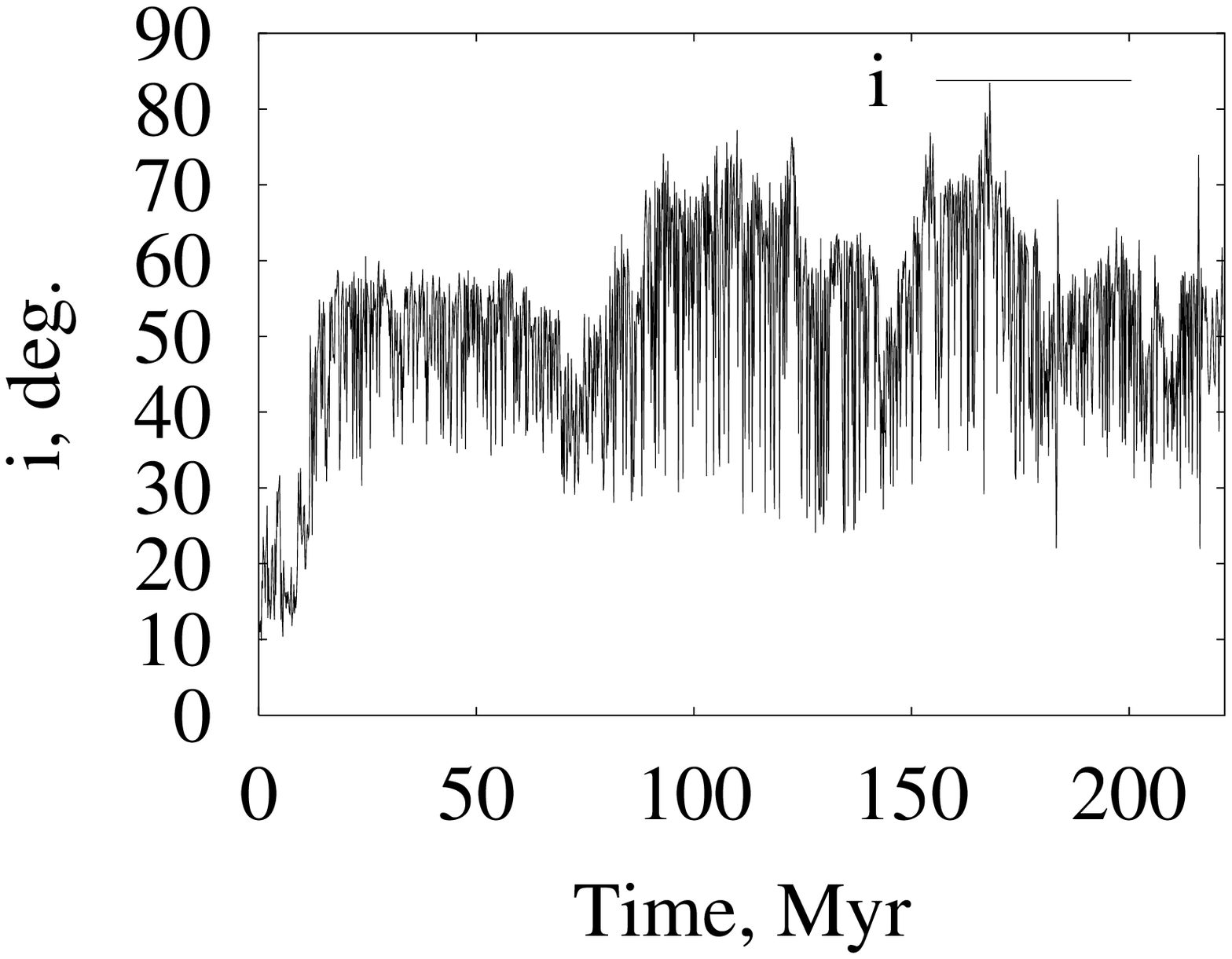}

\includegraphics[width=52mm]{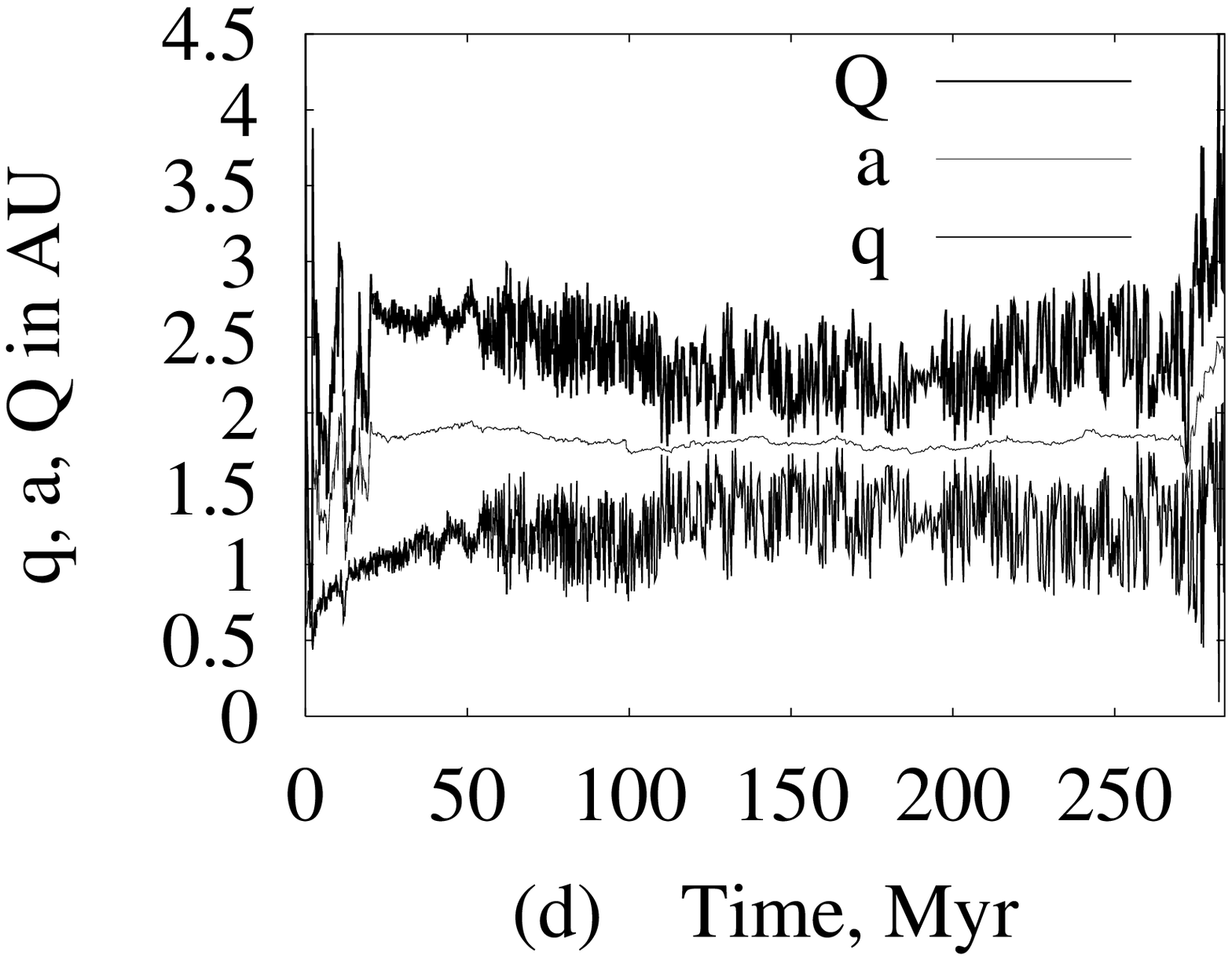}
\includegraphics[width=52mm]{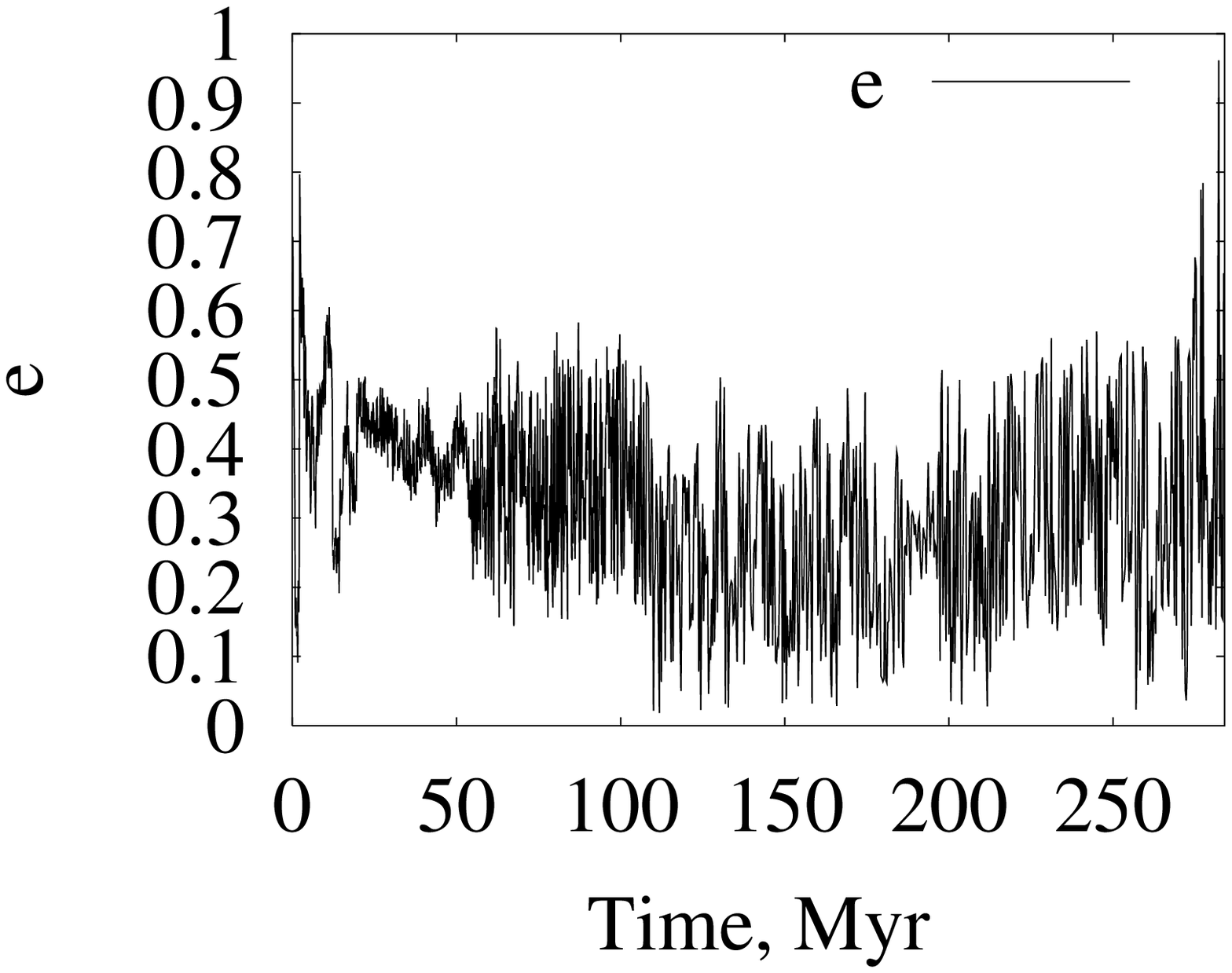}
\includegraphics[width=52mm]{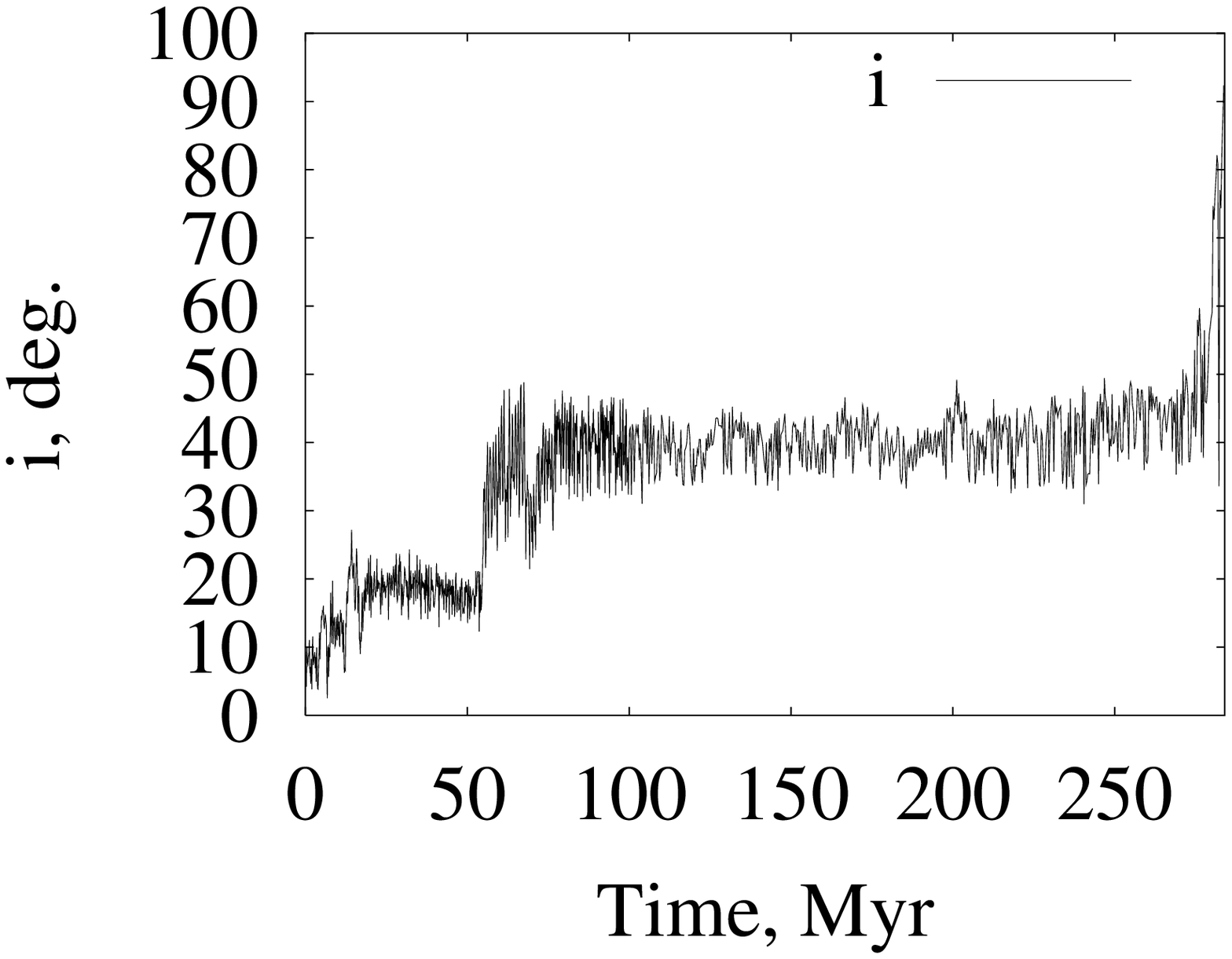}

\includegraphics[width=52mm]{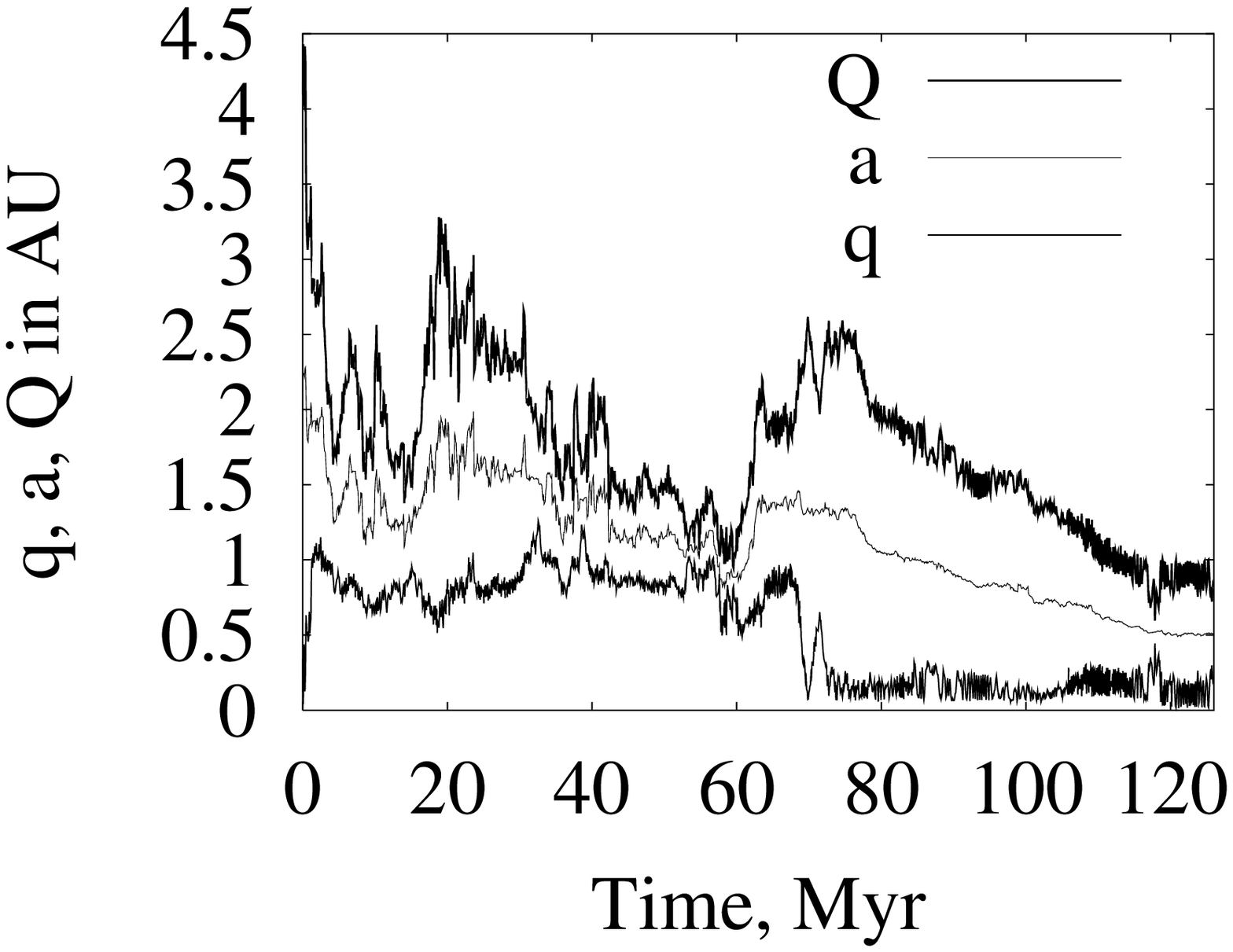}
\includegraphics[width=52mm]{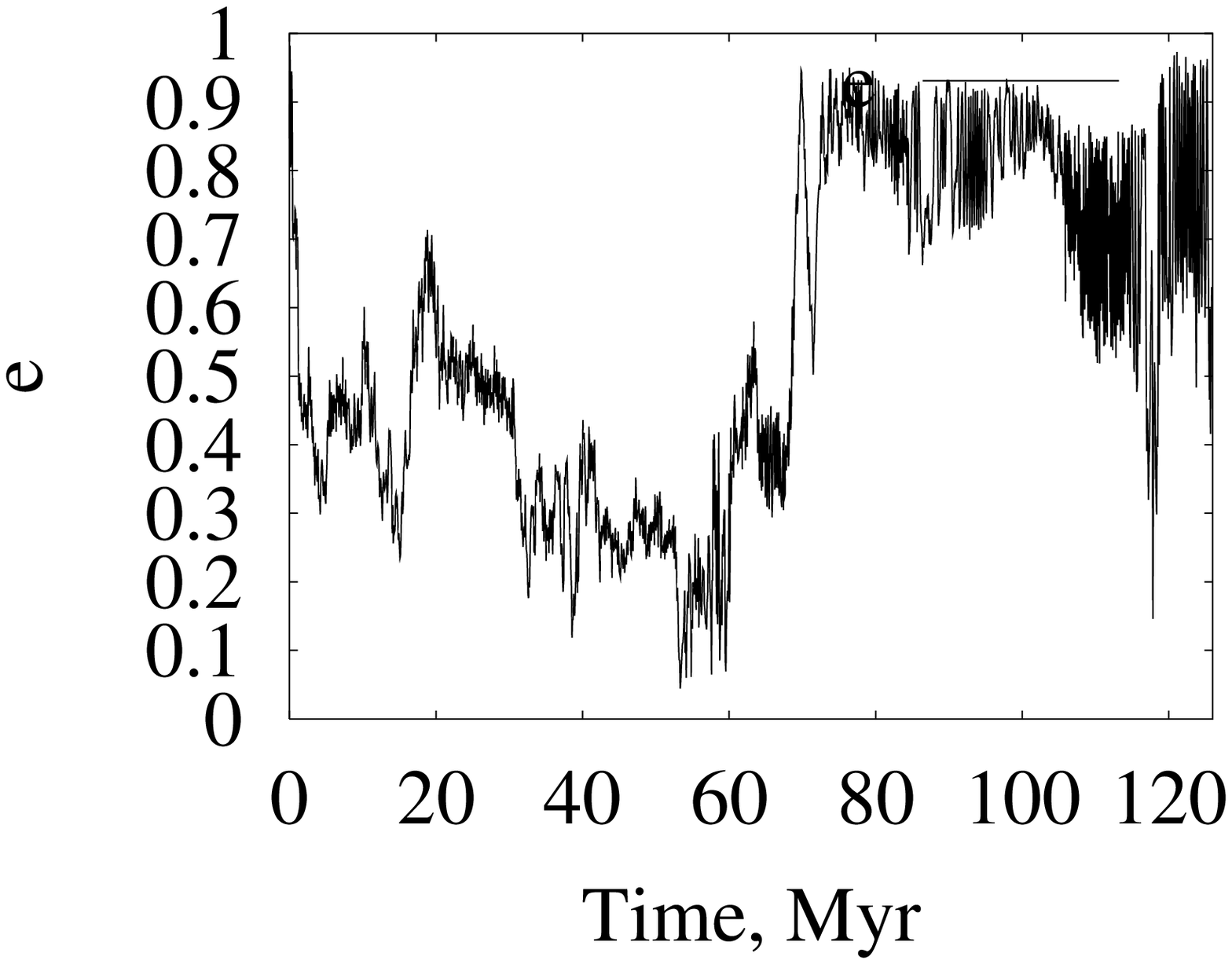}
\includegraphics[width=52mm]{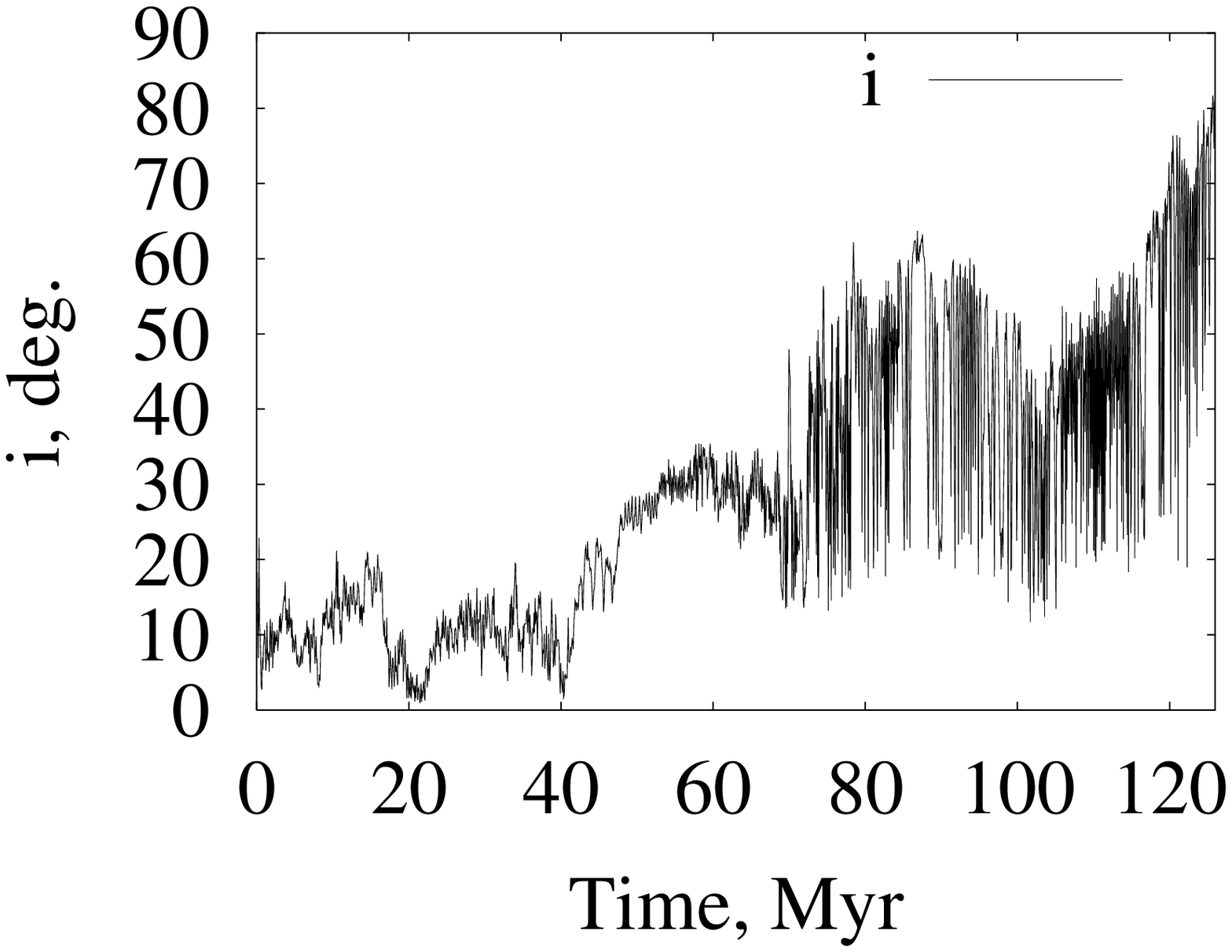}

\caption{Time variations in $a$, $e$, $q$, $Q$, 
% sin($i$) 
and $i$ for a former JCO 
in initial orbit close to that of Comet 10P (a), 2P 
%(b). 
(b, e), 9P (d), or an asteroid
at the 3/1 resonance with Jupiter (c). 
For (a) at $t$$<$$0.123$ Myr $Q$$>$$a$$>$1.5 AU. 
%These two objects are not included in Table 2. 
Results from 
BULSTO code with $\varepsilon \sim 10^{-9}-10^{-8}$ 
(a-c) and by a symplectic method 
with $d_s$=30 days (d) and with $d_s$=10 days (e).
%at $d_s$=10 days (c).
}

\end{figure}%

     In Fig. 3 we present the time in Myr during which objects 
had semi-major axes in an interval with a width of 0.005 AU 
(Figs. 3a-b) or 0.1 AU (Figs. 3c). At 3.3 AU (the 
2:1 resonance with Jupiter) there is a gap for asteroids that 
migrated from the 5:2 resonance and for former JCOs (except 2P). 
 
\begin{figure}
\includegraphics[width=160mm]{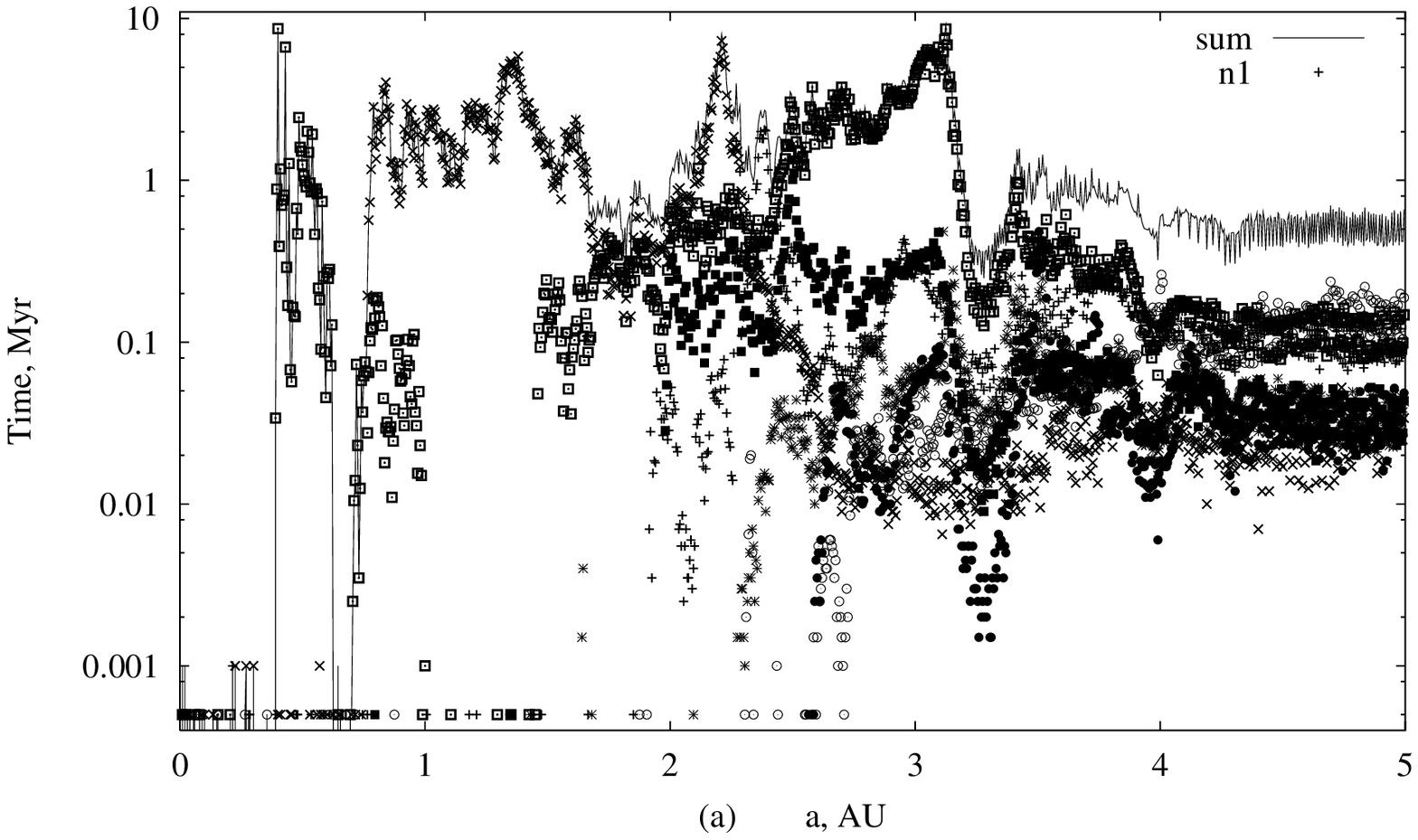}
\includegraphics[width=81mm]{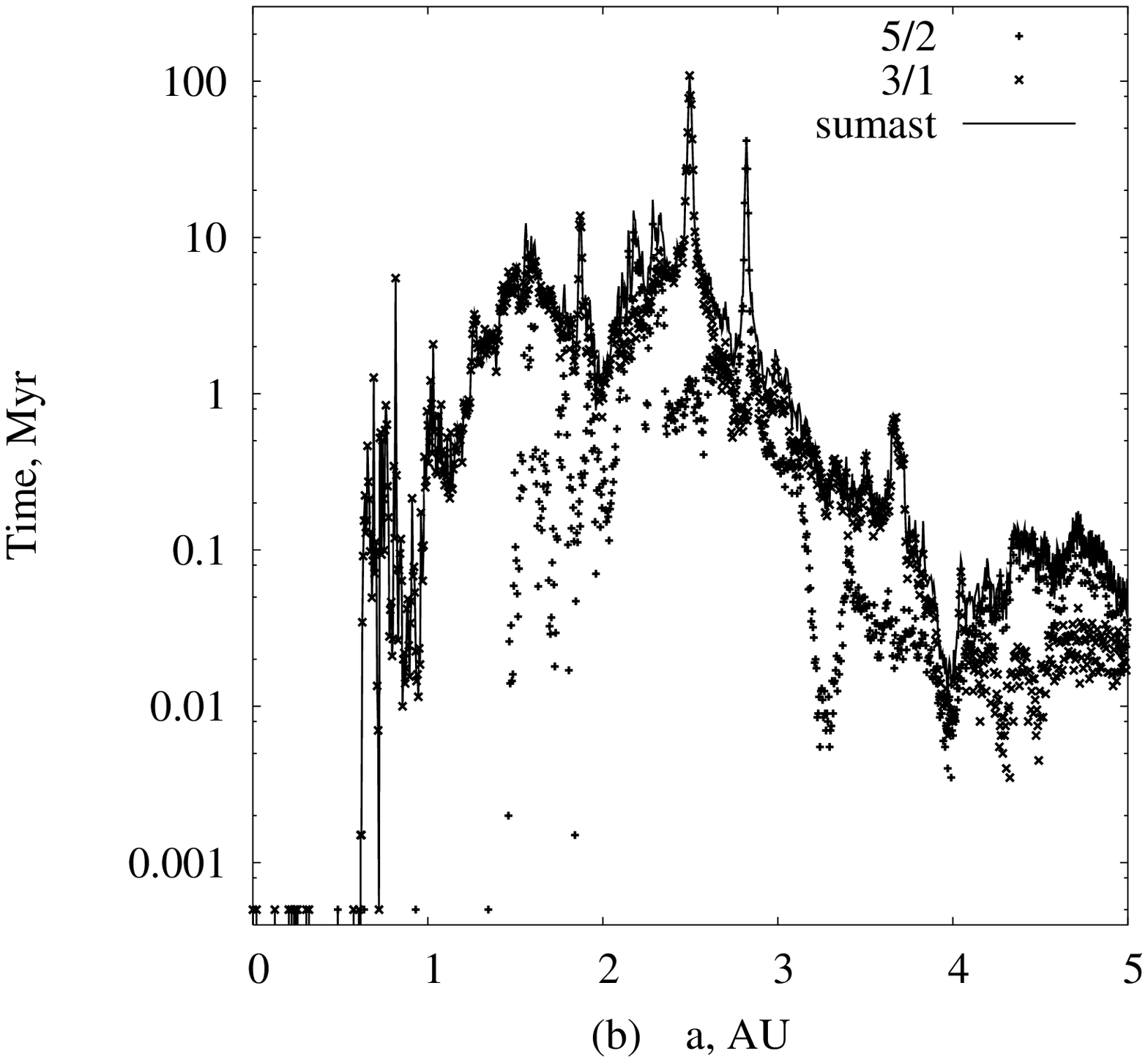} 
\includegraphics[width=81mm]{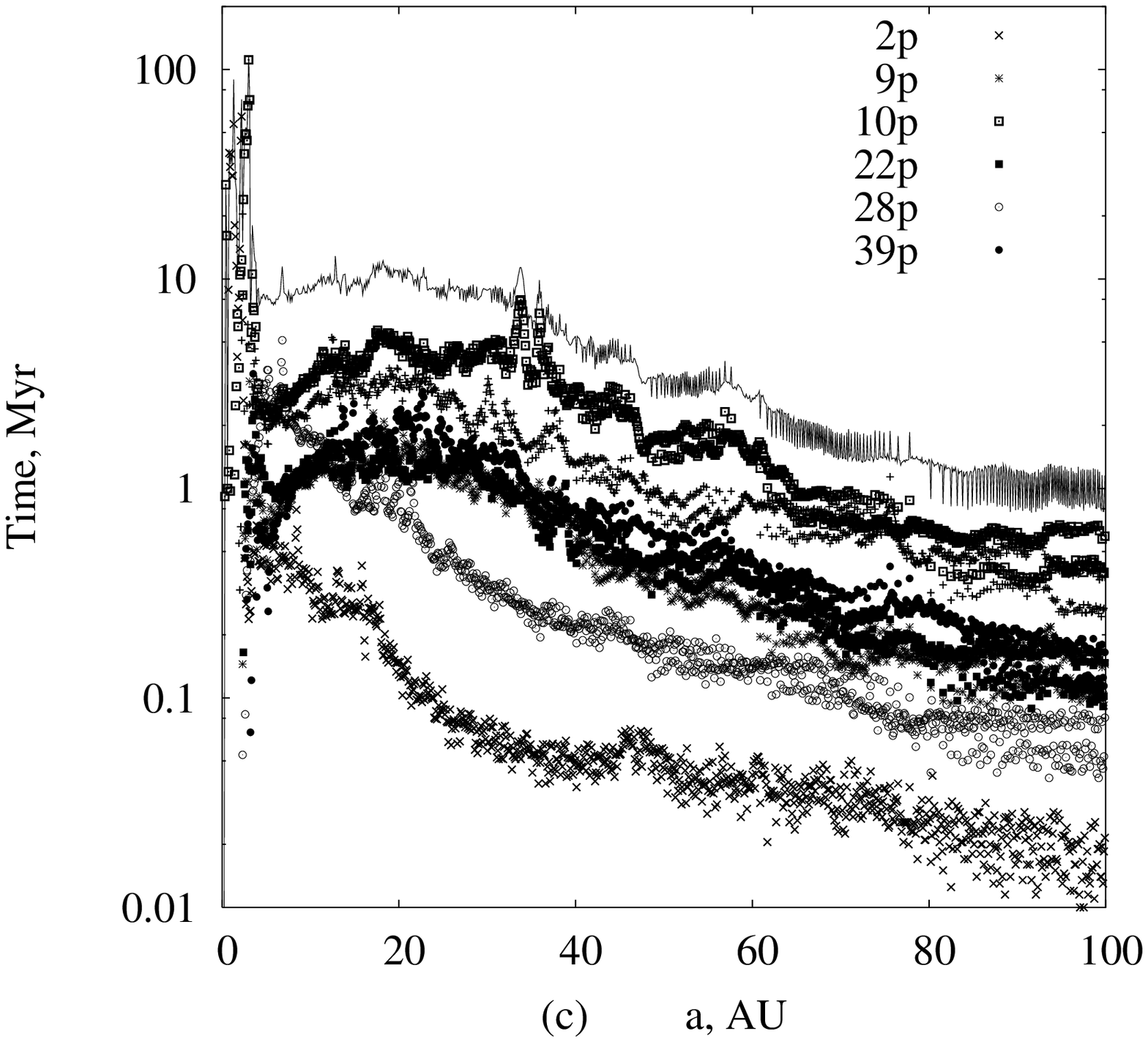}
\caption{Distribution of 7852 migrating JCOs (a, c) 
and 288 resonant asteroids at $e_o$=0.15 and $i_o$=10$^o$
(b) with their semi-major axes.
The curves plotted in (c) at {\it a}=40 AU  are (top-to-bottom) for sum,
10P, n1, 39P, 22P, 9P, 28P, and 2P
(series n2 and 44P are not included in the figure). 
For Figs. (a) and (c), designations are the
same. Results from BULSTO code with $\varepsilon \sim 10^{-9} - 10^{-8}$.} 
\end{figure}%

%\begin{figure}

%\includegraphics[width=60mm]{a197a}
%\includegraphics[width=60mm]{a197e}
%\includegraphics[width=60mm]{a197i}

%\caption{Time variations in $a$, $e$, $q$, $Q$, 
%sin($i$) 
%and $i$
%for a former JCO 
%in initial orbit close to that of Comet 2P. 
%Results from 
%a symplectic method at $d_s$=10 days.}

%\end{figure}%

\begin{figure}
\includegraphics[width=81mm]{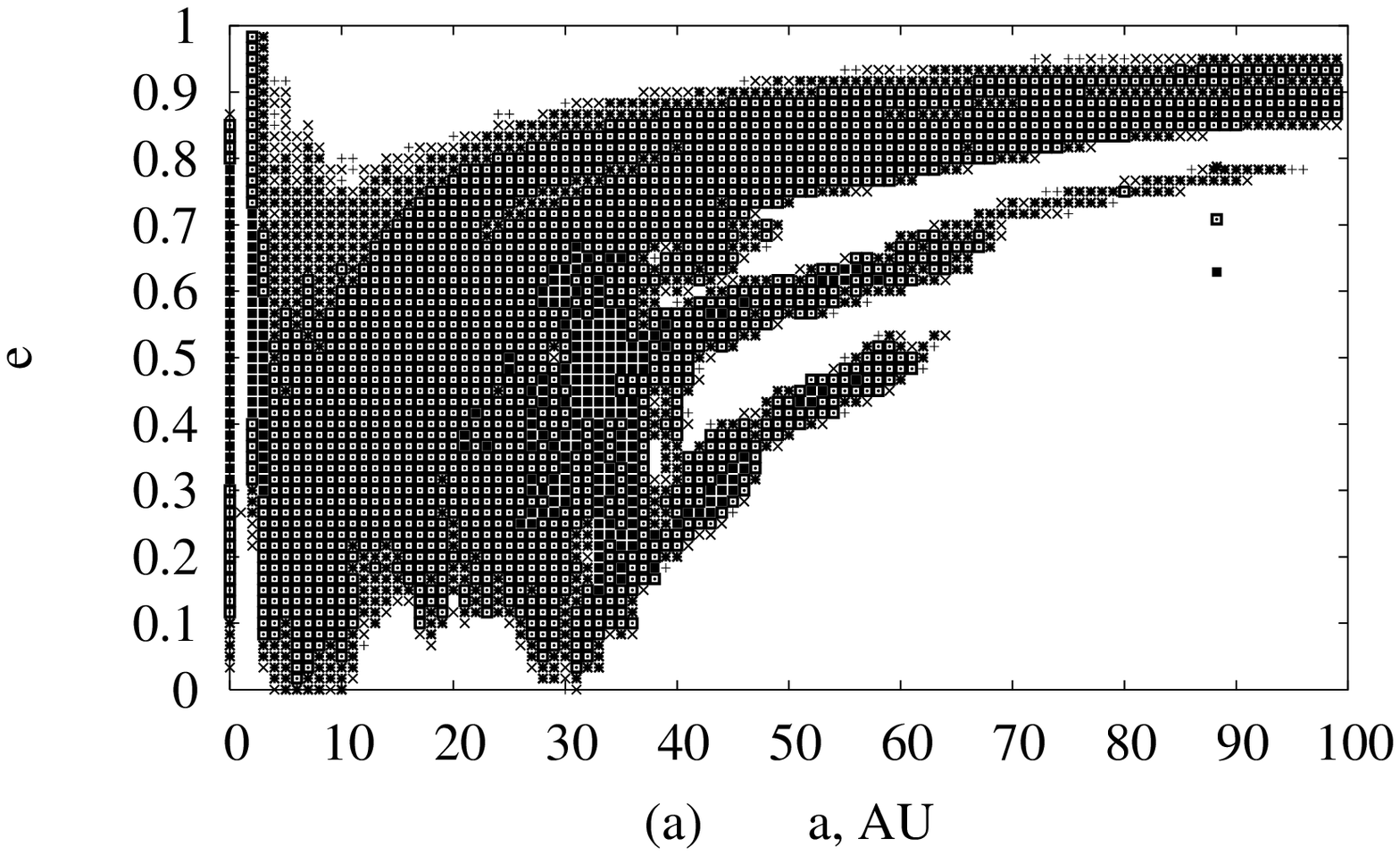}
\includegraphics[width=81mm]{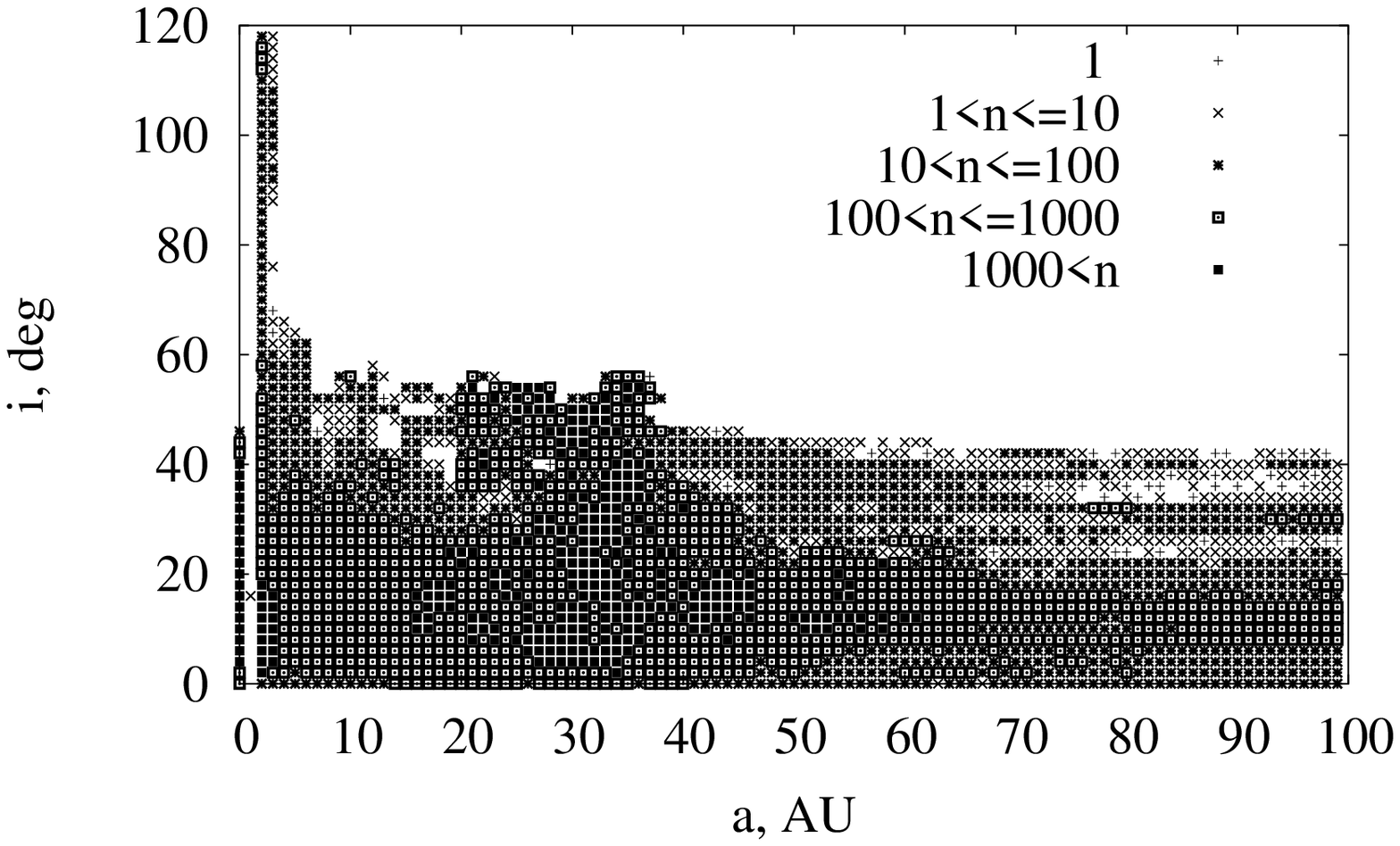}
\includegraphics[width=81mm]{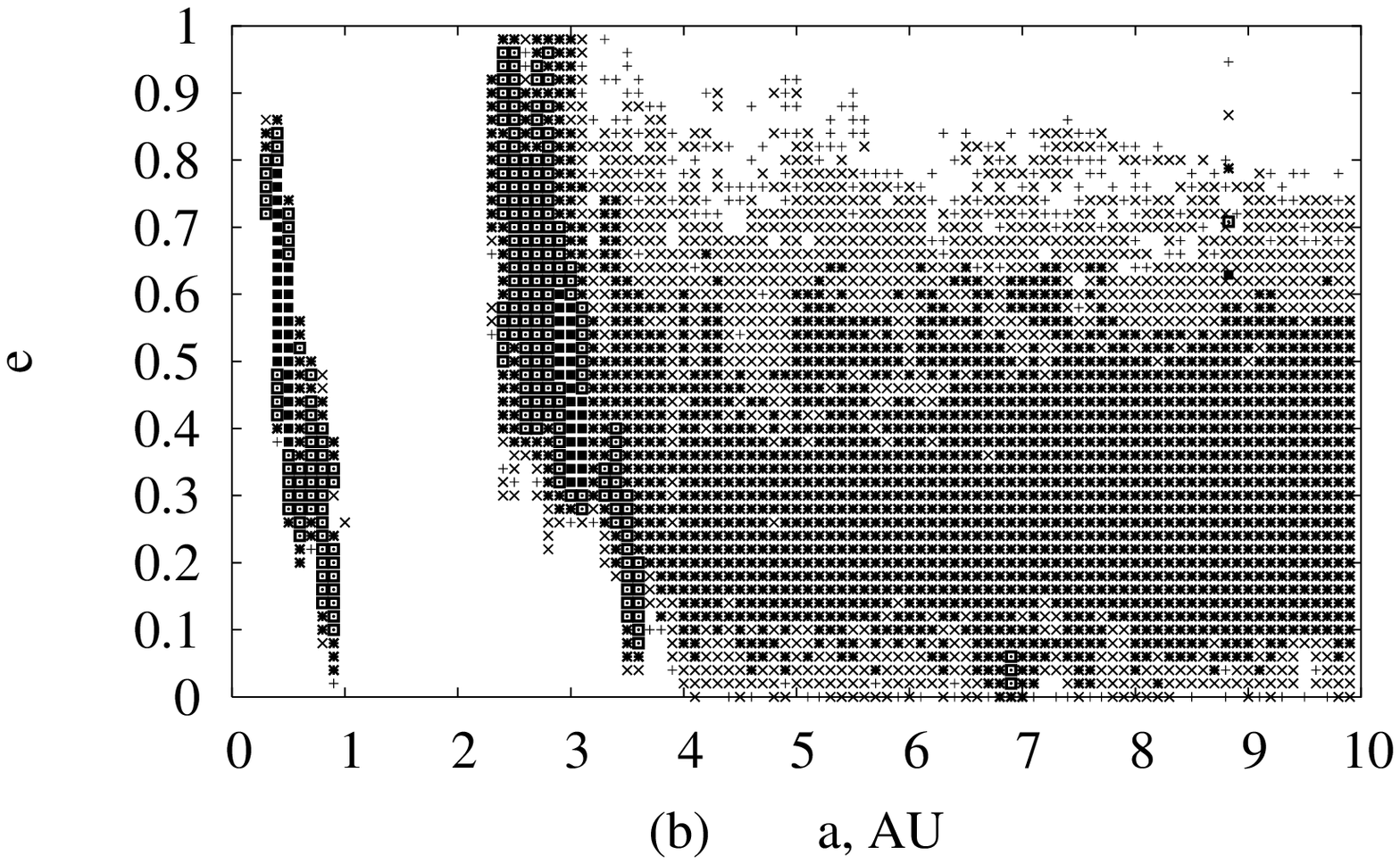}
\includegraphics[width=81mm]{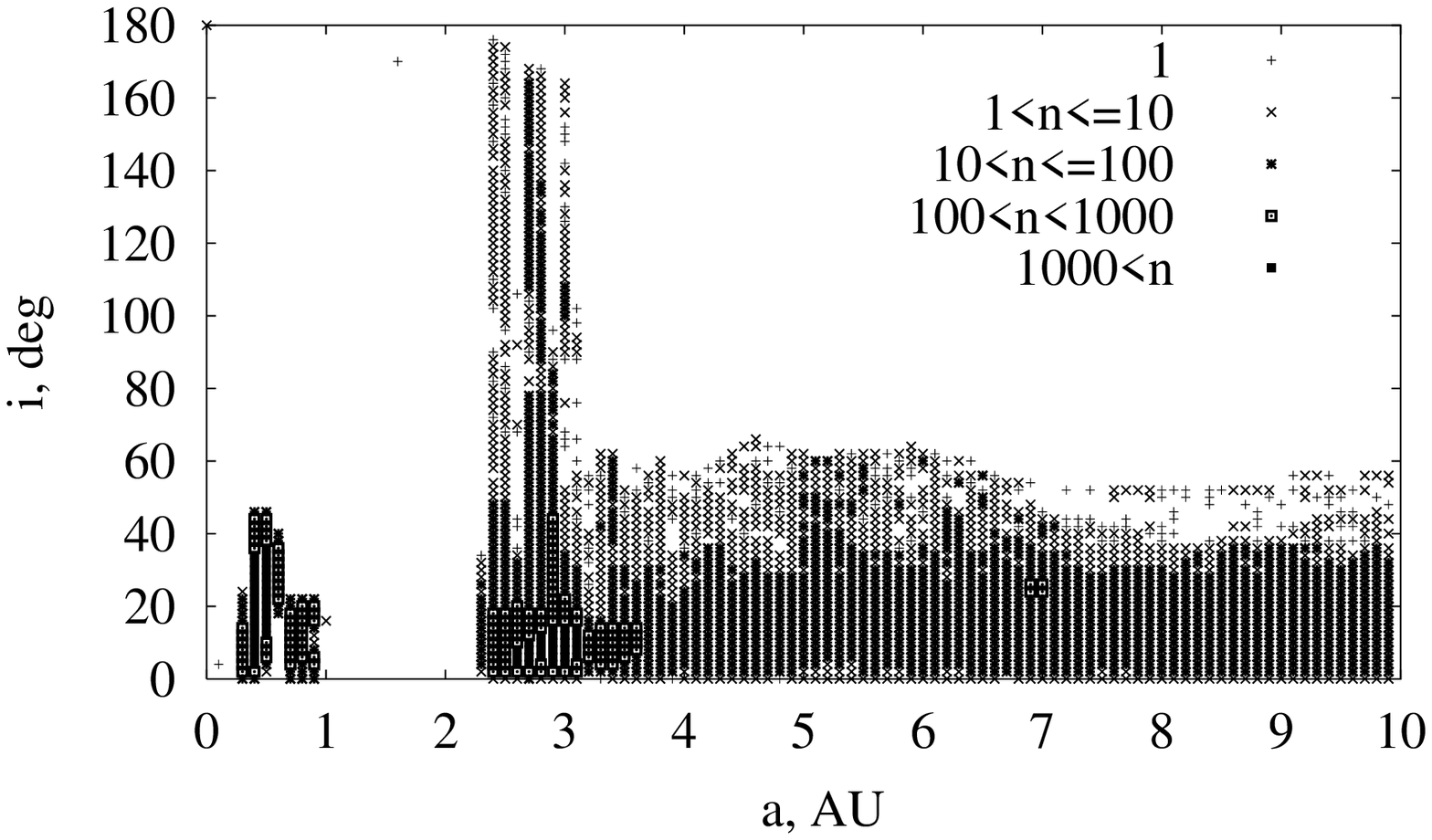}
\includegraphics[width=81mm]{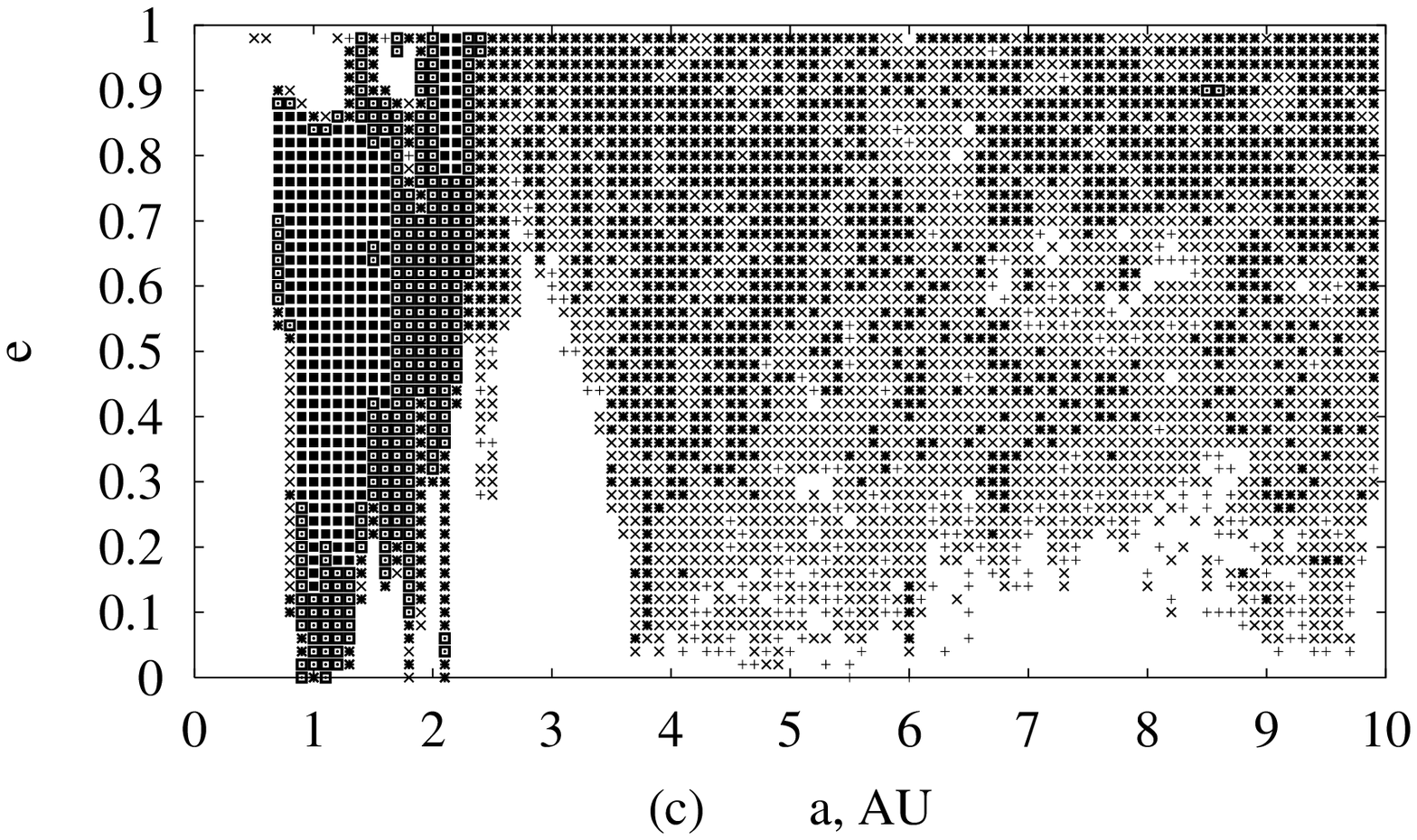}
\includegraphics[width=81mm]{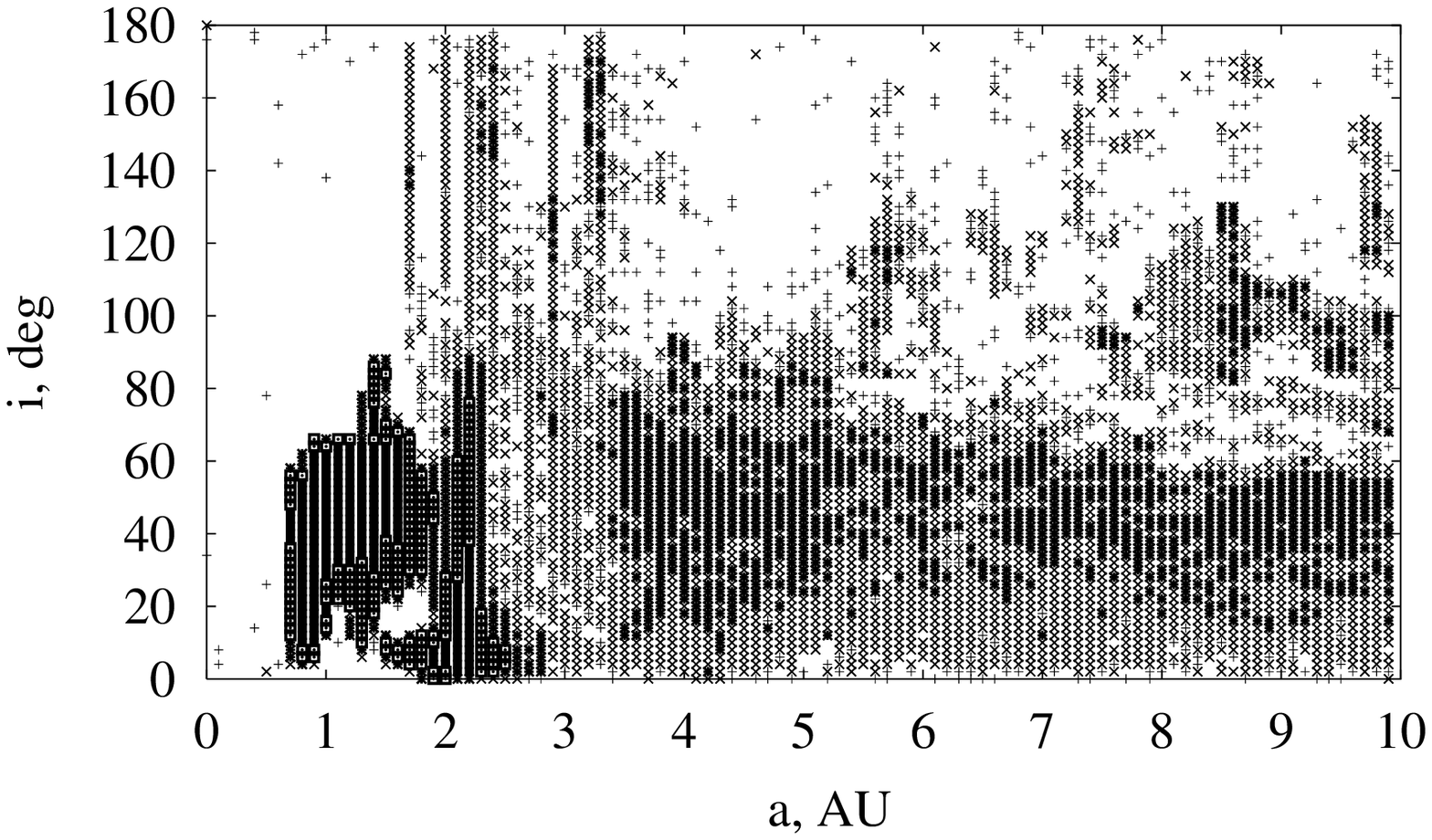}
\includegraphics[width=81mm]{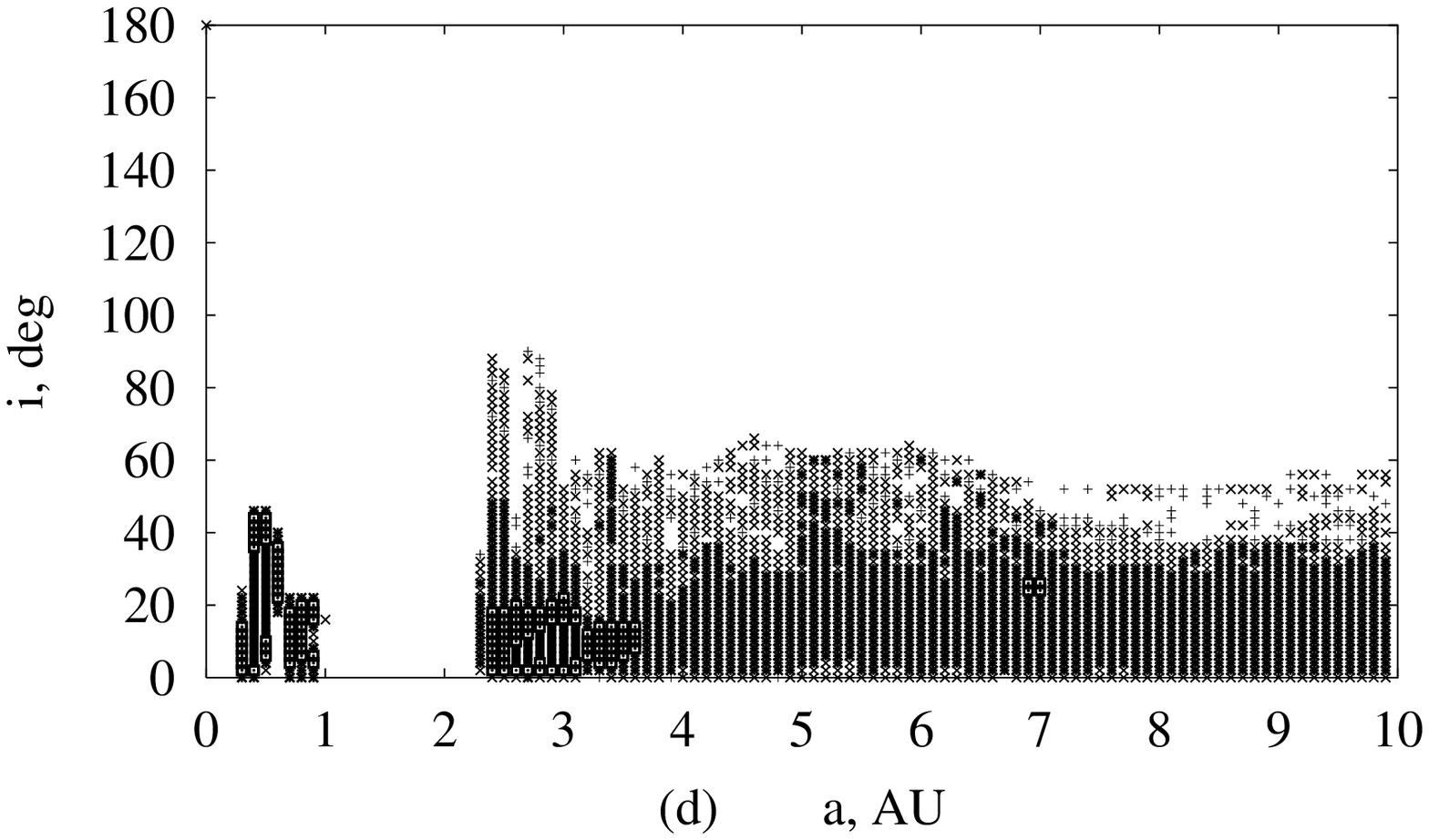}  % new in April
\includegraphics[width=81mm]{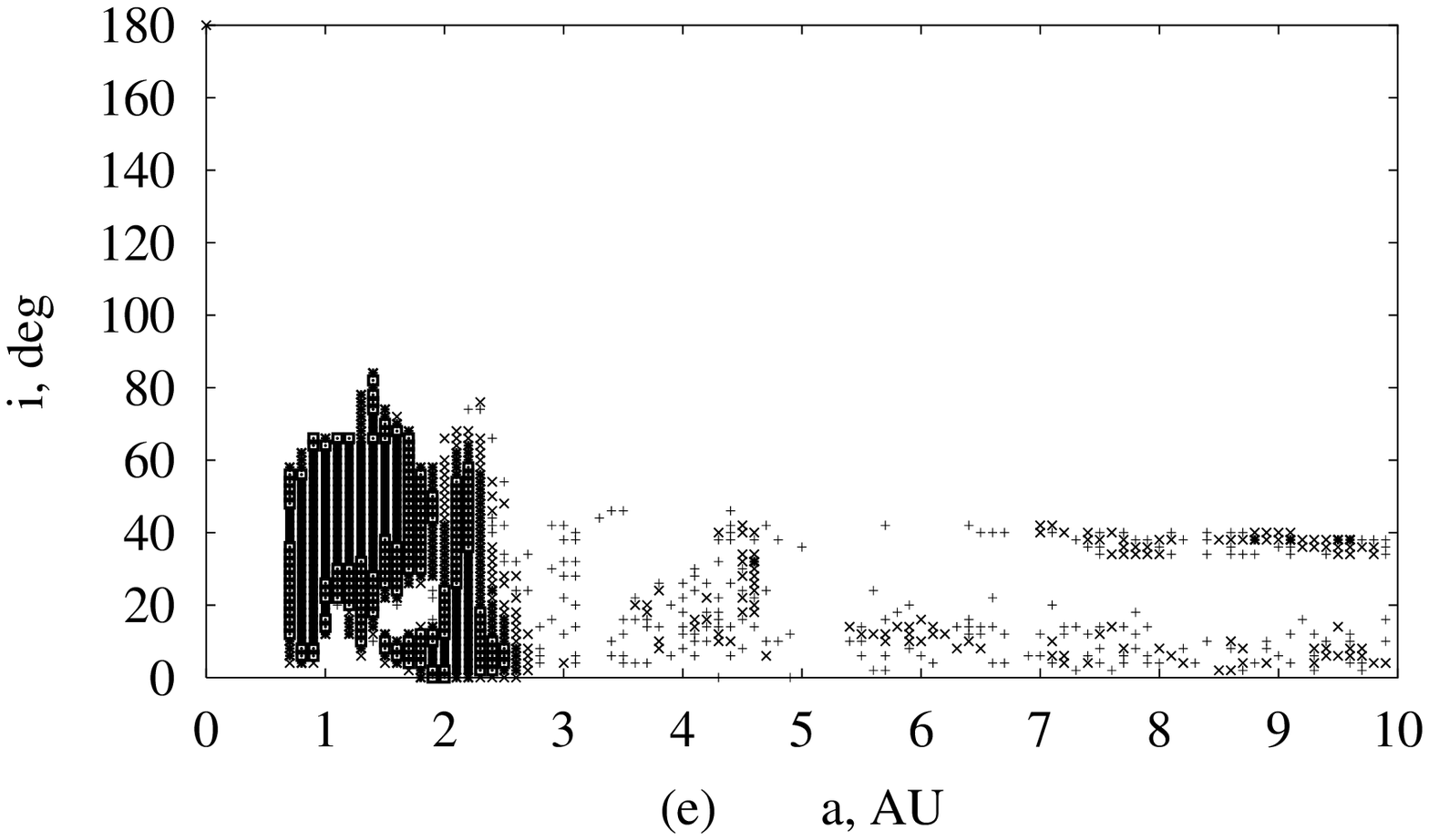}  % new in April

\caption{Distribution of migrating objects in semi-major axes,
eccentricities, and inclinations for objects in initial orbits close to that of 10P (a-b,d),
2P (c,e), and 
%the resonance 3:1 (d).
%%%39P (d). =========
BULSTO code with $\varepsilon \sim 10^{-9} - 10^{-8}$.
For (d-e) it was considered that an object disappeared when perihelion distance became
less than 2 radii of the Sun. In other cases (a-c)
objects disappered when they collided with the Sun. }
\end{figure}%

\begin{figure}
\includegraphics[width=81mm]{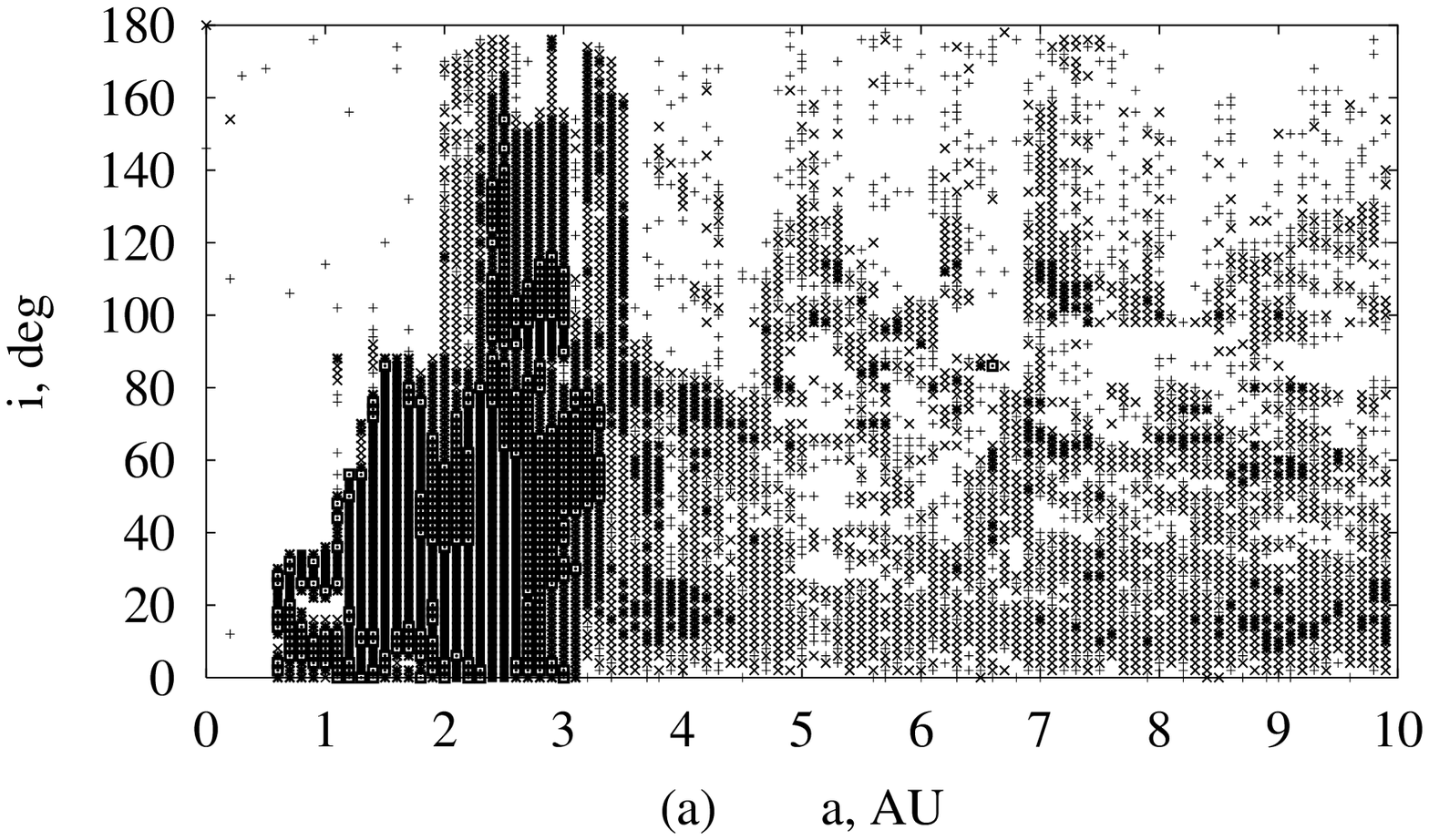}
\includegraphics[width=81mm]{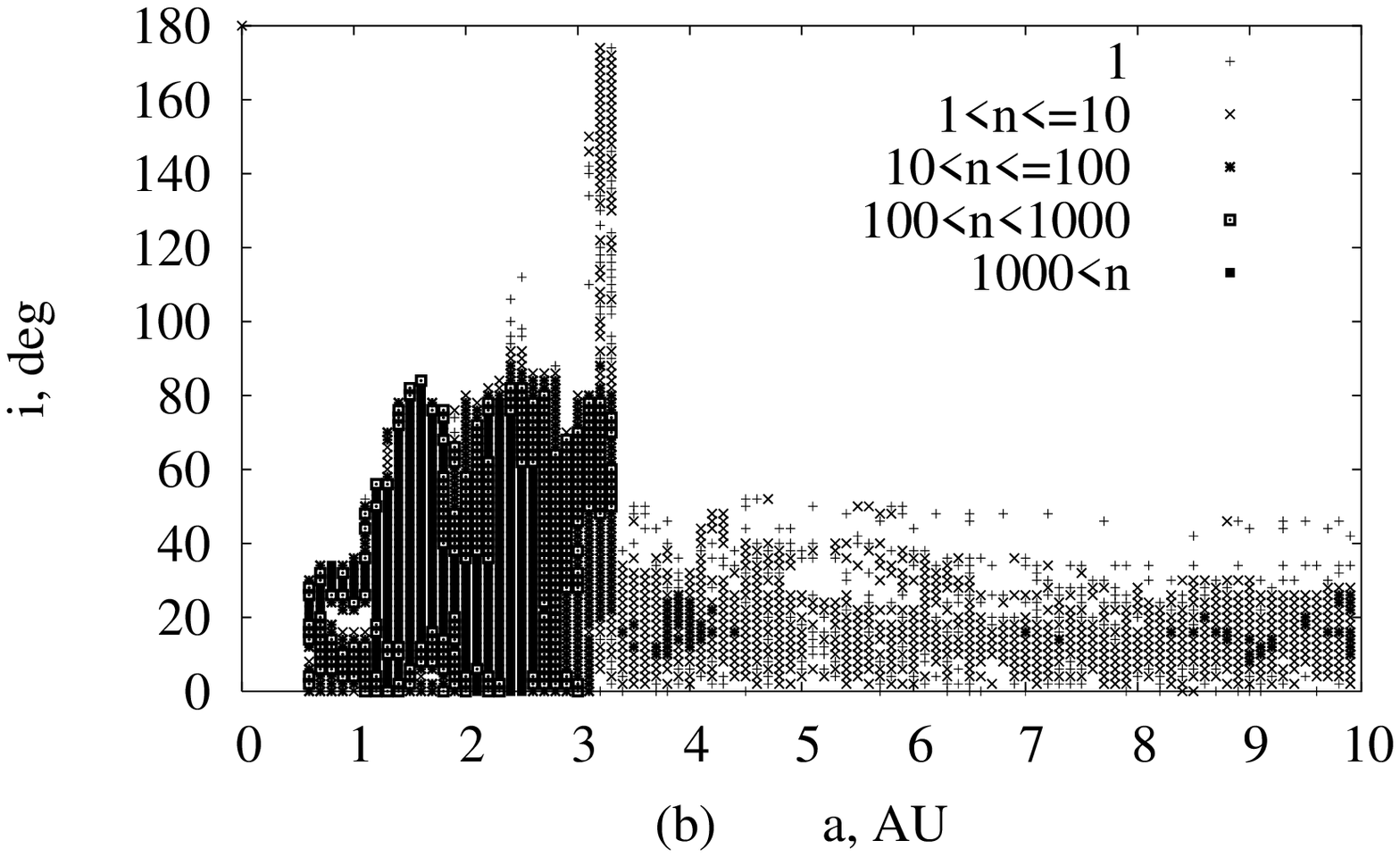}
\caption{Distribution of migrating objects in semi-major axes
and inclinations for objects in initial orbits 
%%close to that of 10P (a), 2P (b), and 
at the 3:1 resonance with Jupiter with $e_o$=0.15 and $i_o$=10$^\circ$.
%39P (d). 
BULSTO code with $\varepsilon \sim 10^{-9} - 10^{-8}$.
For (a) objects disappered when they collided with the Sun. 
For (b) it was considered that an object disappeared when perihelion distance became
less than 2 radii of the Sun. 
} 
\end{figure}%

      For the $n1$ data set, $T_J$=0.12 Myr and, while moving in 
Jupiter-crossing orbits, objects had orbital periods $P_a$$<$$10$, 
10$<$$P_a$$<$20, 20$<$$P_a$$<$50, 50$<$$P_a$$<$200 yr for 11\%, 21\%, 
21\%, and 17\% of $T_J$, respectively. Therefore, there are three 
times as many  JCOs as  Jupiter-family comets (for which $P_a$$<$20 
yr).
      We also found that some JCOs, after residing in orbits with 
aphelia deep inside Jupiter's orbit,
transfer for tens of Myr to the trans-Neptunian region, either in low 
or high eccentricity orbits. We conclude that some of the main belt 
asteroids may reach typical TNO orbits, and then become 
scattered-disk objects having high eccentricities, and vice versa. 
The fraction of objects from the 5:2 resonance that collided with the 
Earth was only 1/6 of that for the 3:1 resonance. Only a small 
fraction  of the asteroids
from the 5:2 resonance reached $a$$<$2 AU (Fig. 3b).

The distributions of migrating former JCOs (2P and 10P)
and resonant asteroids 
in $a$ and $e$ (left) and in $a$ and $i$ (right) are presented in Fig. 4-5. 
For each picture we considered 250 migrating objects
(288 for Fig. 5), 100 intervals for $a$, and about the same number 
of intervals for $e$ and $i$. 
Different designations correspond to different numbers $n$ of orbital elements 
(calculated with a step of 500 yr)  in one bin (in Fig. 4 $`$$<$=$`$ means $\le$). 
%%%%=====Ipatov and Mather (2003) considered 
Similar plots for 39P runs were presented in [17]. 
All the former JCOs  reached low eccentricity orbits 
very rarely with 2$<$$a$$<$3.5 AU and 11$<$$a$$<$28 AU. 
There were many positions of objects when their perihelia were close to a 
semi-major axis of a giant planet, mainly of Jupiter (Fig. 4a). 
Note that Ozernoy et al. [22] considered the migration of Neptune-crossers and 
found that the main concentrations of perihelia were near Neptune's orbit. 
%The pictures are different for different runs. 

%For initial orbits close to that of Comet 2P, 
%orbits often reached $90^\circ$$<$$i$$<$$180^\circ$ with various values 
%of $a$$>$1.7 AU (Fig. 4c). 
%For Fig. 4b (Comet 10P) the values $90^\circ$$<$$i$$<$$120^\circ$ 
%were obtained only at 2.4$<$$a$$<$3 AU.
W. Bottke pointed out that H. Levison showed that
it is difficult to detect solar collisions in any numerical integrator,
so he removed objects with $q$$<$$q_{\min}$. The results presented above
were obtained considering collisions with the Sun, but we also
investigated what happens if we consider $q_{\min}$ equal to $k_S$
radii $r_S$ of the Sun. For $k_S$=2, some results are
presented in Fig. 4d-e, 5b. The only difference with the runs
that considered collisions with the Sun is that  for those runs
for series 2P and 10P and for the 3:1 resonance, 
some objects reached $90^\circ$$<$$i$$<$$180^\circ$ (mainly with 
2$<$$a$$<$3.5 AU) (Fig. 4b-c, 5a). For $k_S$=2 there were no comets with  
$i$$>$90$^\circ$
and there were only a few orbits of asteroids with $i$$>$90$^\circ$ 
(Fig. 4d-e, 5b). The consideration of $q_{\min}$ at $k_S$=3 did not
influence the collision probabilities with the terrestrial planets or
getting orbits with $a$$<$2 AU. For example, with BULSTO for the two objects with the 
largest collision probabilities, the time spent in orbits with
$a$$<$2 AU decreased by only 0.3\% for 2P at $k_S$=3 and was the same
for 10P at $k_S$=10. 

\section*{COMPARISON OF ORBIT INTEGRATORS}

To determine the effect of the choice of orbit integrators and 
convergence criteria, we made additional runs with BULSTO 
at $\varepsilon$=10$^{-13}$ and $\varepsilon$=10$^{-12}$ 
and with a symplectic integrator. 
The orbital evolution of 9551 JCOs was computed with the RMVS3 code. 
%For BULSTO we used $\varepsilon$=10$^{-13}$ 
%and $\varepsilon$=10$^{-12}$ instead of the $10^{-8} - 10^{-9}$ range 
%used in the previous section.  
For the  symplectic method we usually used an 
integration step $d_s$ of 3, 10, and 30 days.
For series $n2$ and 44P we took different values of $d_s$ between
5 and 10 days.

We find that, % the results are not the same.
%, and conclude that the choice does make a significant difference.  
%On the other hand, 
exclusive for the case of close encounters
with the Sun, the differences between integrator choices 
(with $d_s$$\le$10 days) are comparable to the differences between 
runs with slightly different initial conditions.  Our interpretation 
is that 1) very small numbers of particles contribute most of the 
collision probabilities
with the terrestrial planets, 2) runs with larger numbers of particles are 
more reliable, and 3) small differences in initial conditions or in 
the errors of the orbit integrators modify the trajectories 
substantially, especially for those particles making major changes in 
their orbits due to close encounters or resonances.  We conclude that 
for the purposes of this paper, the various choices of orbit 
integrator are sufficiently equivalent.

To illustrate these points, Tables 3-4  present the results obtained by
BULSTO with $\varepsilon$$\le$10$^{-12}$ and the symplectic method 
with $d_s$$\le$10 days.
Most of the results obtained with these values of $\varepsilon$ and 
$d_s$ are statistically
similar to those obtained for $10^{-9} $$\le$$ \varepsilon$$\le$$ 
10^{-8}$. For example,
a few objects spent millions of years in Earth-crossing orbits 
inside Jupiter's orbit (Figs. 1-2), and their probabilities of collisions with 
the Earth were thousands of times greater than for more typical 
objects.
For series $n1$ with $d_s$$\le$10$^d$, the
probability of a collision with Earth for 
one object with initial orbit close to that of Comet 44P was 88.3\%
of the total probability for 1200 objects from this series, and the 
total probability for 1198 objects was only 4\%.
This object and the object presented in Fig. 2e and in the third line of Table 5
were not included in Table 3 with $N$=1199 for $n1$ and with $N$=250 for 2P, 
respectively.
For the 3:1 resonance  with $d_s$=10 days, 142 objects spent 140 and 
84.5 Myr in IEO and Aten orbits, respectively, even longer than
for $\varepsilon$$\sim$$10^{-9}$-$10^{-8}$.
Additionally, up to 40 Myr and 20 Myr were spent in such orbits by 
two other objects which had estimated probabilities of collisions 
with the terrestrial planets greater than 1. 
For the 2P runs with $\varepsilon$$\le$$10^{-12}$ and $N$=100, the 
calculated objects spent 5.4 Myr in Apollo orbits with $a$$<$2 AU.

%3/1 step 10 tx=190.46-50.46=140 my for 142 objects; aten 108.9-24.2=84.5

The values of $P_r$   are usually of the same order of
magnitude for different methods (see Tables 3-4), and the difference between the data 
is comparable to the differences between different runs belonging to a series.
For Earth and Venus, the values of $P_r$  are
about 1--4 for Comets 9P, 22P, 28P and 39P; for Comet 44P they are about 4--6.  
For 28P and 39P with the symplectic method
$P_r$ is about twice that for BULSTO. For 10P the values of $P_r$ are about 20--40 and
are several times larger than  for the above series, and 
for 2P they exceed 100 and are several times larger than for 
10P. With the $n1$  and $n2$ runs, for Earth $P_r$$>$4 and $P_r$$>$10, respectively.
The ratio of $P_r$ to the mass of the planet was typically several 
times larger for Mars than for Earth and Venus. The main difference 
in $P_r$ was found for the 3:1 resonance.
In this case greater values of $P_r$ were obtained for $d_s$=10 days
than for BULSTO. 
As noted above, a few exceptional objects dominated  the 
probabilities, and 
for the 3:1 resonance two objects, which had collision probabilities 
greater than unity for the terrestrial planets, were not included 
in Table 4.  These two objects can  increase the total value of $P_r$ 
for Earth by a factor of several.
%Time variations of orbital elements for Comet 2P obtained at $d_s$=10 days are 
%presented in Fig. 2e.

The mean time $T_d$ (in Kyr) 
spent in orbits with $Q$$<$4.2 AU can differ by three orders of magnitude
for different series of runs (Tables 3-4). 
%In Table 5
%in [ ] we present the number $N_d$ of objects each of which got $Q$$<$4.2 AU
%during at least 1000 yr, the second number $N_d$ in [ ] means the same 
%but only for $q$$<$1.017 AU. For series 2P and the 3:1 resonance with Jupiter
%both values of $N_d$
%the latter two numbers 
%were almost the same, for 10P they differed by a factor of
%2.9 (symplectic runs) or 3.7 (BULSTO), and for other series only a small
%portion of decoupling objects crossed the orbit of the Earth.
For most runs (except for 2P and asteroids) the number of objects
which got $Q$$<$4.7 AU was several times larger than 
that for $Q$$<$4.2 AU. 

%We also did some 
For symplectic runs with $d_s$=30 days for most of the 
objects we got  results similar to those with $d_s$$\le$10 days, 
but about 0.1\%  of the
objects    reached Earth-crossing orbits with $a$$<$2 AU for several 
tens of Myr (e.g., Fig. 2d) 
and even IEO orbits. These few bodies increased the 
mean value of $P$ by a factor of more than 10. With $d_s$=30 days, 
four  objects from the runs $n1$, 9P, 10P had a probability of 
collisions
with the terrestrial planets greater than 1 for each, and for 2P 
there were 21 such objects among 251 considered.
For resonant asteroids, we also obtained much larger values than 
those for BULSTO for  $P$ and $T$ for RMVS3 with 
$d_s$=30 days, and similarly  for the 3:1 resonance even with $d_s$=10 
days. For this resonance it may be
better to use $d_s$$<$10 days.
Probably, the results of symplectic runs with $d_s$=30 days can be considered as 
such migration that includes some nongravitational forces.

      In the case of close encounters with the Sun (Comet 2P and 
resonant asteroids), the probability
$P_S$ of collisions with the Sun  
during lifetimes of objects
was larger for RMVS3 than for BULSTO, and for
$10^{-13} $$\le$$ \varepsilon $$\le$$ 10^{-12}$ it was greater than 
for $10^{-9}$$ \le$$ \varepsilon $$\le$$ 10^{-8}$
($P_S$=0.75 for the 3:1 resonance % 3:1
with $d_s$=3 days). This probability is presented in Table 7 for several runs.

\begin{table}[h]

\begin{center}
\caption{Probability of collisions with the Sun (for asteroids $e_o$=0.15 and $i_o$=10$^o$).}
$ \begin{array}{lllllll}
\hline
  & \varepsilon=10^{-13} & \varepsilon=10^{-12} & \varepsilon=10^{-9} 
& \varepsilon=10^{-8} & d_s=10 $ days$ & d_s=30 $ days$ \\
\hline
$Comet 2P$& 0.88 & 0.88 & 0.38 & 0.32& 0.99 & 0.8 \\

$resonance $ 3:1 &0.46& 0.5& 0.156 & 0.112 & 0.741 & 0.50 \\

$resonance $ 5:2 & &0.06& 0.062 & 0.028 & 0.099 & 0.155\\
\hline
\end{array} $

\end{center}
\end{table}

For Comet 2P the values of $T_J$ were much smaller for RMVS3 than 
those for BULSTO and they were smaller
for smaller $\varepsilon$; for other runs these values do not depend 
much on the method.
In our opinion, the most reliable values of  $T_J$ were 
obtained with $10^{-13} $$\le$$ \varepsilon$$ \le$$ 10^{-12}$.
In the direct integrations reported by Valsecchi et al. [23], 13 of 
the 21 objects fell into the Sun,
so their value of $P_S$=0.62 is in accordance with our results 
obtained by BULSTO; it is less than
that for $\varepsilon$=$10^{-12}$, but greater than  for 
$\varepsilon$=$10^{-9}$.
Note that even for different $P_S$ the data presented in Tables 2-3 
usually are similar. As we did not calculate
collision probabilities of objects with planets by direct 
integrations, but instead calculated them with the random phase 
approximation from the
orbital elements, we need not make integrations with extremely high 
accuracy. We showed [24] that for BULSTO
the integrals of motion were conserved better and the plots of 
orbital elements for closely separated values of  $\varepsilon$ were
closer to one another with  $10^{-9} $$\le$$ \varepsilon $$\le$$ 
10^{-8}$. The smaller the value of $\varepsilon$, the more
integrations steps are required, so $\varepsilon$$\le$$10^{-12}$ for 
large time intervals are not necessarily better than those for 
$10^{-9} $$\le$$ \varepsilon $$\le$$ 10^{-8}$. Small $\varepsilon$ is 
clearly necessary for close encounters. 
%Ipatov and Hahn (1999) and Ipatov (2000) 
Therefore we
made most of our  BULSTO runs with  $10^{-9} $$\le$$ \varepsilon 
$$\le$$ 10^{-8}$.
We found [1],[9] that former JCOs
reached resonances more often for BULSTO than for RMVS3 with 
$d_s$=30 days.
For a symplectic method it is better to use smaller $d_s$ at a 
smaller distance $R$ from the Sun,
but in some runs  $R$ can vary considerably during the evolution.
The choice of $d_s$ depends on the smallest values of $R$, 
so a symplectic method with a constant $d_s$
may not be effective when $R$ is very different
for different objects considered.

%For RMVS3, former JCOs got resonant orbits less often than for BULSTO and more
%often collided with the Sun. For example, for Comet 2P Encke 
%$P_S$=0.99 at $d_s$=10 days instead of $P_S$=0.35 for BULSTO.
%Using a symplectic method, Levison and Duncan (1994) showed that 
%Comet 2P became sungrazing at $t$=0.05 Myr,
%and at the direct integrations by Valsecchi et al. (1995) 13 of the 
%21 such objects fell into the Sun.

\section*{MIGRATION FROM BEYOND JUPITER TO THE EARTH}

The fraction $P_{TNJ}$ of TNOs reaching
Jupiter's orbit under the influence of the giant planets in 1 Gyr is
0.8-1.7\% [6]. As the mutual gravitational influence of TNOs can play a 
lar\-ger role in variations of their orbital elements than collisions [2],
%(Ipatov, 2001), 
we considered the upper value of $P_{TNJ}$.
Using the total of $5\times10^9$ 1-km 
TNOs with $30$$<$$a$$<$$50$ AU,  and assuming
that the mean time for a body to move in a Jupiter-crossing orbit is 
0.12 Myr, we find that about $N_{Jo}$=$10^4$ 1-km former TNOs are now 
Jupiter-crossers, and 3000 are Jupiter-family comets.
With the total times spent by $N_J$ simulated JCOs 
in Apollo orbits we can estimate the number of 1-km former TNOs now moving 
in such orbits using the following formula: 
$N_{Apollo}=N_{Jo} \cdot t_{Apollo}/(N_J\cdot t_J)$, where 
$t_{Apollo}$ is the total time during which $N_J$ former JCOs moved in Apollo orbits, 
and $N_J\cdot t_J$ is the total time during which $N_J$ JCOs moved in 
Jupiter-crossing orbits. 
Similar formulas can be considered for other types of orbits. Based on $n1$ and $n2$ 
series of runs, we obtain that there are about 700-900 Amors and 2000-3000 Apollos 
(the last numbers include very eccentric orbits) which came from the Edgeworth-Kuiper 
belt and have diameters greater than 1 km.
Even if the number of Apollo objects is an order of magnitude smaller
than the above value, it may still be comparable to the real number 
(750) of 1-km
Earth-crossing objects (half of them are in orbits with $a$$<$2 AU), 
although the
latter number does not include those in highly eccentric orbits.

The ratio $k_2$ of the number of Apollos with $a$$<$2 AU to the 
number of all Apollos was very different for different series 
of runs. It was 0.4 for $n1$ runs, but the total time spent by 
3100 considered objects in orbits with $a$$<$2 AU was due mainly 
only to one object. For $n2$ series with RMVS3, $k_2$=0.2, but 
practically all time spent by 6250 objects in orbits with $a$$<$2 AU 
was also due to one object. For all JCOs considered, $k_2$ was about 0.5, 
but the total time spent in orbits with $a$$<$2 AU was due mainly to the 
objects from 2P series of runs.

The values of the characteristic time (usually $T_c$) for the 
collision of a former JCO or a resonant
asteroid with a planet (see Tables 3-4) are greater than the values 
of $T_f$ for NEOs in Table 1,
%(for the Earth $T_c=T/P\approx2.7$ Gyr for 7850 considered JCOs and 
%%%$T_c$$\approx$1.1 Gyr for 7852 objects,  
%while for observed Earth-crossing objects it is $\approx$100 Myr),
so we expect that the mean inclinations and eccentricities
of unobserved NEOs are greater than those for the NEOs that are 
already known.  Similar results were found in [25].
% because our migrating JCOs and resonant asteroids moved in more 
%inclined and eccentric orbits than observed NEOs.
On average, the values of $T_c$ for our $n1$ and $n2$ series  and for most of 
our simulated JCOs were not greater than those for our calculated 
asteroids, and migrating Earth-crossing objects had similar $e$ and 
$i$ for  both former JCOs and resonant asteroids.
Former JCOs, which move in Earth-crossing orbits for more than 1 Myr, 
while moving in such orbits, usually had larger $P$ and smaller $e$ and $i$
(sometimes similar to those of the observed NEOs, see Figs. 1-2).
It is easier to observe orbits with smaller values of $e$ and $i$, and 
probably, many of the NEOs moving in orbits with large values of $e$
%(especially with $e$$\approx$1) 
and $i$ have not yet been discovered.
About 1\% of the observed Apollos cross Jupiter's orbit, and an 
additional 1\% of Apollos have aphelia between 4.7-4.8 AU,
but these Jupiter-crossers are far from the Earth  most of time, so 
their actual fraction of ECOs is greater
than  for observed ECOs. The fraction of Earth-crossers among 
observed Jupiter-family comets is about 10\%.
This is a little more than $T/T_J$ for our $n1$ runs, but less than 
for $n2$ runs.
For our former resonant asteroids, $T_J$ is relatively large 
($\approx$0.2 Myr), and such asteroids can reach cometary orbits.

Comets are estimated to be active for $T_{act}$$\sim$$10^3$--$10^4$ yr. 
$T_{act}$ is smaller for closer encounters with the Sun [5], %(Weissman et al., 2002),
so for Comet 2P it is smaller than for other JFCs.
Some former comets can move for tens or
even hundreds of Myr in NEO orbits,  so the number of extinct comets 
can exceed the number of active comets by several orders of magnitude.
The mean time spent by Encke-type objects in Earth-crossing orbits is 
$\ge$0.4 Myr (even for $q_{\min}$). % at $k_S$=2). 
This time corresponds to $\ge$40-400  
%(or may be even larger number) 
extinct comets of 
this type. Note that the diameter of Comet 2P %Encke
is about 5-10 km, so the number of smaller extinct comets can be much larger.

      The above estimates of the number of NEOs are approximate. For 
example, it is possible that the number of 1-km TNOs is several times 
smaller than $5\times10^9$,
while some scientists estimated that this number can be up to $10^{11}$ [13].
%(Jewitt, 1999). 
Also, the fraction of TNOs that have migrated towards the Earth might be 
smaller. On the other hand, the above
number of TNOs was estimated for $a$ $<$ 50 AU, and TNOs from more 
distant regions can also migrate inward. 
Probably, the Oort cloud  could also supply Jupiter-family comets. 
According to Asher et al. [26], the rate of a cometary
object decoupling from the Jupiter vicinity and transferring to an 
NEO-like orbit is increased by a
factor of 4 or 5 due to nongravitational effects
(see also [27]).
%Fernandez and Gallardo, 2002). 
This would result 
in  larger values of $P_r$ and $T$ than those shown in Table 3.
%We consider that collisions of JCOs with asteroids could increase 
%the number of
%objects migrating to the near-Earth space, especially to orbits with $a<2$ AU.
%     The less are the masses of objects, the more often they can 
%change their orbits due to collisions with smaller bodies.
%     Considering collision probability of an objects with radius of 
%1 km with a smaller object
%     to be $10^{-18}$ and $10^9$ impactors (probably, it is the 
%number of asteroids with $d$=25 m),
%     we obtain that the probability of such collision during 0.1 Myr 
%is 0.0001.

      Our estimates show that, in principle, the trans-Neptunian belt 
can provide a significant portion %(up to several tens percent)
of the Earth-crossing objects, 
although many NEOs clearly came from the main asteroid belt. 
Many
former Jupiter-family comets can have orbits typical of asteroids, 
and collide with the Earth from typical NEO orbits.
It may 
be possible to explore former TNOs near the Earth's orbit without 
sending spacecraft to the trans-Neptunian region.

%      More than half of the close encounters of active comets with the 
%Earth belong to long-period comets, which amount to about 80\% of the 
%known population. Thus, though probabilities are smaller for larger 
%eccentricities, the number of collisions of both long-period and 
%short-period active comets with the inner planets can be of the same 
%order of magnitude. According to our results, 

      Based on the estimated collision probability $P=4\times10^{-6}$ 
we find that  1-km former TNOs now collide with the Earth once in 
3 Myr. This value of $P$ is smaller
than that for our $n1$, and especially than for $n2$, 10P and 2P runs.
Assuming the total mass of planetesimals that ever crossed Jupiter's 
orbit is $\sim$$ 100m_\oplus$, where
$m_\oplus$ is the mass of the Earth [1],[28],
%(Ipatov, 1993, 2000), 
we conclude that the total mass of bodies that impacted the
Earth is $4\times10^{-4} m_\oplus$. If ices comprised only half of this mass,
then the total mass of ices $M_{ice}$ that were delivered to the 
Earth from the feeding zone of
the giant planets is about the mass of the terrestrial oceans 
($\sim2\times10^{-4} m_\oplus$).

      The calculated probabilities of collisions of objects with 
planets show that the fraction of the mass of the planet  delivered 
by short-period comets can be greater for Mars and Venus than for the 
Earth 
%(compare the values of $P$$/$$m_{pl}$ using $P$ from 
(Table 3). 
This larger mass fraction would result in 
relatively large ancient oceans on Mars and Venus. On the other hand, 
there is the deuterium/hydrogen paradox of Earth's oceans, as the D/H 
ratio is different for oceans and comets.  Pavlov et al. [29] %(1999) 
suggested that solar wind-implanted hydrogen on interplanetary dust 
particles could provide the necessary low-D/H component of Earth's 
water inventory, and Delsemme [30] %(1999) 
considered that most of the seawater 
was brought by the comets that originated in Jupiter's zone, 
where steam from the inner solar system condensed onto icy 
interstellar grains before they accreted into larger bodies.

Our estimate of the migration of water to the early Earth is in 
accordance with [31], but %Chyba (1989), but
is greater than those of Morbidelli et al. [32] and Levison et al. [33].
%Levison et al. (2001) concluded
%that the Earth has accreted only 5\% of its oceans from bodies that migrated
%from the feeding zones of Uranus and Neptune.
%Morbidelli et al. (2000) and Levison et al. (2001)
The latter obtained smaller
values of $M_{ice}$, and we suspect that this is because they did not 
take into account the migration of bodies into orbits with $Q$$<$4.5 AU.
Perhaps this was
because they modeled a relatively small number of objects, and 
Levison et al. [33] did not take into
account the influence of the terrestrial planets. In our runs the 
probability of a collision of a single object
with a terrestrial planet could be much greater than the total 
probability of thousands of other objects, so the statistics are 
dominated by rare occurrences that might not appear in smaller 
simulations.
The mean probabilities of collisions can  differ by orders of 
magnitude for different JCOs.
Other scientists considered other initial objects and smaller numbers 
of Jupiter-crossing objects, and
  did not find decoupling from Jupiter, which is a rare event.
We believe there is no contradiction between our present results and 
the smaller
migration of former JCOs to the near-Earth space that was obtained in 
earlier work,
including our own papers (e.g. [9]),
%Ipatov and Hahn, 1999), 
where we used the same integration package.

%There is additional evidence from the classification of objects based 
%on their albedos. 
From measured albedos, Fernandez et al. [34] %(2001) 
concluded that
the fraction of extinct comets among NEOs and unusual asteroids is significant
(at least 9\% are candidates).
The idea that there may be many extinct comets among NEOs was 
considered by several scientists.
Rickman et al. [35] believed that comets
played an important and perhaps even dominant role among all km-size 
Earth impactors.
In their opinion, dark spectral classes that might include the 
ex-comets are severely underrepresented
(see also [10]). %Jewitt and Fernandez, 2001).
Our runs showed that if one observes
former comets in NEO orbits, then it is  probable that they have already
moved in such orbits for millions (or at least hundreds of thousands) 
years, and only a few of them have been
in such orbits for short times (a few thousand years).
Some former comets that have moved in typical NEO orbits for millions 
or even hundreds of millions of years,
and might have had multiple close encounters with the Sun
(some of these encounters can be very close to the Sun, e.g.
in the case of Comet 2P at $t$$>$0.05 Myr), could have lost their mantles,
which causes their low albedo, and so change their albedo
(for most
%  97\% 
observed NEOs, the albedo is greater than that for comets [34]) %Fernandez et al., 2001)
and would look like typical asteroids 
or some of them could disintegrate into mini-comets.
Typical comets have larger rotation periods than typical NEOs [36]-[37],
%(Binzel et al., 1992, Lupishko and Lupishko, 2001),
but, while losing considerable portions of their masses, extinct 
comets can decrease these periods. 
%In future we plan to consider a 
%larger initial number of objects initially located beyond Jupiter
%in order to better estimate their
%probabilities of migration to a near-Earth space.
For better estimates of the portion of extinct comets among NEOs
 we will need orbit
integrations for many more TNOs and JCOs, and wider analysis of 
observations and craters.  

\section*{CONCLUSIONS}

%Collision statistics for the terrestrial planets
%are dominated by very small numbers of bodies 
%that reach orbits with high collision probabilities, so it is 
%essential to consider very large numbers of particles. The initial 
%conditions for the orbit integrations appear to matter more than the 
%choice of orbit integrator.  
Some Jupiter-family comets can reach 
typical NEO orbits and remain there for millions of years. While the 
probability of such events is small
(about $\sim$0.1\%), % in our runs and perhaps smaller for other initial data), 
nevertheless the majority of collisions of former JCOs with the 
terrestrial planets are due to such objects.  
Most former TNOs that have typical NEO orbits moved in such
orbits for millions of years 
(if they did not disintegrate into mini-comets), so during most of this time
there were extinct comets. From the dynamical point of view there 
could be %(not `must be`) 
many extinct comets among the NEOs. The amount of water 
delivered to the Earth during planet formation could be about the 
mass of the Earth oceans.

\section*{ACKNOWLEDGMENTS}

      This work was supported by NRC (0158730), NASA (NAG5-10776), 
INTAS (00-240), and RFBR (01-02-17540).
      For preparing some data for figures we used some subroutines 
written by P. Taylor.
We are thankful to W. F. Bottke, S. Chesley, and H. F. Levison for 
helpful discussions.

\centerline{}

\end{document}